                                                                  \documentclass[12pt,preprint]{aastex}
\shorttitle{Sunspot rotation, filament, and flare: The event on 2000 February 10}
\shortauthors{Yan et al.}

\begin{document}

%\title{Two successive M-class flares associated with two coronal mass ejections triggered by the collision of two small sunspots with opposite magnetic polarities and %movement directions}
\title{Sunspot rotation, sigmoidal filament, flare, and coronal mass ejection: The event on 2000 February 10}
\author{X. L. Yan\altaffilmark{1,2}, Z. Q. Qu\altaffilmark{1}, D. F. Kong\altaffilmark{1,2}, C.L. Xu\altaffilmark{3}}

\altaffiltext{1}{National Astronomical Observatories/Yunnan
Astronomical Observatory, Chinese Academy of
              Sciences, Kunming, Yunnan, 650011, China.}
\altaffiltext{2}{Graduate School of Chinese Academy of Sciences,
           Zhongguancun, Beijing, 100049, China.}
\altaffiltext{3}{Yunnan Normal University, Kunming, Yunnan, 650092, China.}
\begin{abstract}
We find that a sunspot with positive polarity had an obvious counter-clockwise rotation and resulted in the formation and eruption of an inverse S-shaped filament in NOAA active region (AR) 08858 from 2000 February 9 to 10. The sunspot had two umbrae which rotated around each other by 195 degrees within about twenty-four hours. The average rotation rate was nearly 8 degrees per hour. The fastest rotation in the photosphere took place during 14:00UT to 22:01UT on February 9, with the rotation rate of nearly 16 degrees per hour. The fastest rotation in the chromosphere and the corona took place during 15:28UT to 19:00UT on February 9, with the rotation rate of nearly 20 degrees per hour. Interestingly, the rapid increase of the positive magnetic flux just occurred during the fastest rotation of the rotating sunspot, the bright loop-shaped structure and the filament. During the sunspot rotation, the inverse S-shaped filament gradually formed in the EUV filament channel. The filament experienced two eruptions. In the first eruption, the filament rose quickly and then the filament loops carrying the cool and the hot material were seen to spiral into the sunspot counterclockwise. About ten minutes later, the filament became active and finally erupted. The filament eruption was accompanied with a C-class flare and a halo coronal mass ejection (CME). These results provide evidence that sunspot rotation plays an important role in the formation and eruption of the sigmoidal active-region filament.
\end{abstract}


\section{Introduction}
Sunspot rotational motions have been observed by many authors
for many decades (Evershed, 1910; Maltby, 1964;
Gopasyuk, 1965). Stenflo (1969) and Barnes
\& Sturrock (1972) suggested that the rotational motion of a
sunspot may be involved with energy build-up and the build-up energy is released by a flare later.

With the high spatial and temporal resolution of recent satellite-borne telescopes,
the observations of rotating sunspots are easily obtained (Nightingale et al. 2002). Using white-light images from TRACE, Brown et al. (2003) analyzed the
rotation speed of the umbrae and penumbrae of several rotating
sunspots. They found that the average rotation speed of the penumbrae of the rotating sunspots was larger than that of the umbra of the rotating sunspots. Through the method of time-distance helioseismology, Zhao \& Kosovichev (2003) found the evidence of structural twist beneath the visible
surface of a rotating sunspot. The rotating sunspots related to other magnetic structures were also identified by many authors. R$\acute{e}$gnier \& Canfield (2006) found that the slow rotation of the sunspot in NOAA AR 8210 enabled the storage of magnetic energy and allowed for
the release of magnetic energy as C-class flares. Tian \& Alexander (2006) found that the sunspot and the sunspot group exhibited a counterclockwise rotation. The twist of the active-region magnetic fields was dominantly left handed. The vertical current and the current helicity
were predominantly negative. Later, Yan \& Qu (2007) presented that sunspot rotation resulted in the appearance of the $\Omega$ magnetic loop in the corona and finally the $\Omega$ magnetic loop erupted as a M-class flare. Zhang,
Li \& Song (2007) reported that a flare was caused by the interaction
between a fast rotating sunspot and ephemeral regions. In addition,
Schrijver et al. (2008) used non-linear force-free modelling to show
the evolution of the coronal field associated with a rotating sunspot,
and suggested that the flare energy comes from an emerging twisted
flux rope. The detailed information about
the polarities, rotation directions and helicities of rotating sunspots in cycle 23 was presented by Yan, Qu \& Xu (2008).
The active regions with rotating sunspots were classified into six types by Yan, Qu \& Kong (2008). They also found that several types have higher flare productivity.
Using multi-wavelength observations of Hinode, Yan et al. (2009) and Min \& Chae (2009) studied the rapid rotation of a sunspot in NOAA active region 10930 in detail. They found extraordinary counterclockwise rotation of the sunspot with positive polarity before an X3.4 flare. Moreover, the sheared loops and an inverse S-shaped magnetic loop in the corona formed gradually after the sunspot rotation. From a series of vector
magnetograms, Yan et al. (2009) found that magnetic force lines are highly sheared along the neutral line accompanying the sunspot rotation. Through analyzing the buildup of the energy and the helicity associated with the eruptive flare on 2005 May 13, Kazachenko et al. (2009) found that sunspot rotation alone can store sufficient energy to power a very large flare. Sunspot rotation may be the primary driver of helicity production and injection into
the corona (Zhang, Flyer, \& Low 2006; Zhang, Liu, \& Zhang 2008; Kumar, Manoharan, \& Uddin 2010; Park et al. 2010; Ravindra, Yoshimura, \& Dasso 2011).

The sigmoid structure wes often observed to be precursors to CMEs (Sterling \& Hudson 1997; Sterling et al. 2000; Pevtsov 2002; Liu et al. 2007; Jiang et al. 2007; Green \& kliem 2009; Bi et al. 2011) and statistically more likely to erupt (Hudson et al. 1998; Canfield, Hudson \& McKenzie 1999; Canfield et al. 2007). The eruptions of the sigmoid structures or filaments are usually involved with flares and CMEs (Jing et al. 2004; Wang et al. 2007; Yan, Qu \& Kong 2011). Amari et al. (2000) presented that the shearing motion resulted in the formation of an S-shaped flux rope by MHD simulation. The emergence of the flux tube can also exhibit a sigmoid structure (Magara \& Longcope 2001; Fan 2001; Gibson et al. 2004). A double-J loop pattern can be merged into full S-shaped loops by a slip-running tether-cutting reconnection in the coronal hyperbolic flux tube (Moore et al. 2001; Aulanier et al. 2010). Tripathi et al. (2009) found the coexistence of a pair of J-shaped hot arcs at temperature T $>$ 2 MK with an S-shaped structure at somewhat lower temperature (T $\approx$ 1-1.3 MK). Some observational findings provide strong evidence to support the bald-patch separatrix surface model (Titov \& D$\acute{e}$moulin 1999) for the sigmoid (McKenzie \& Canfield 2008). Other observations and simulations supposed that the X-ray
sigmoid appears at the quasi-separatrix layer between the flux rope and external fields (Gibson et al. 2002; Low \& Berger 2003, Kliem, Titov, \& T$\ddot{o}$r$\ddot{o}$k 2004; Savcheva \& van Ballegooijen 2009). Liu et al. (2002) reported that the sigmoid structure was formed by the reconnection of the emerging flux and the pre-existing field. Green, Kliem, \& Wallace (2011) exhibited that the flux cancellation at the internal polarity inversion line resulted in the formation
of a soft X-ray sigmoid along the inversion line and a coronal mass ejection. By using reconstructed 3D coronal magnetic field, R$\acute{e}$gnier \& Amari (2004) found that the sigmoid was higher than the filament in the corona, while the filament and the sigmoid had the same orientation. Consequently, the formation of the sigmoid structure remains an interesting open question.

In this paper, we present a clear case of the S-shaped active-region filament formation and eruption caused by the sunspot rotation in NOAA active region 08858 on Febuary 10, 2000.

\section{Observations}
The NOAA AR 08858 was observed by several
spacecrafts from 2000 February 9 to 10. The active region was located at N28E01 with $\beta$ field configuration of the sunspot
group on February 9, 2000. This active region was a very productive active region. It produced 13 C-class, 3 M-class and 1 X-class flares during its
journey over the whole solar disk.

The observation of Transition Region and Coronal Explorer (TRACE) covered the whole process of this event from white-light to EUV wavelength. The data of
TRACE white-light, 1600 \AA\ and Fe IX/X 171 \AA\ images have a cadence of about 30 seconds - 1 minute and a pixel size of 0.$^\prime$$^\prime$5 (Handy et al. 1999).
Full-disk line-of-sight magnetograms are used to show the magnetic fields in the photosphere. The magnetograms were taken by the Michelson Doppler Imager (MDI) on
board the Solar and Heliospheric Observatory (SOHO) (Scherrer et al. 1995) with a 96-min cadence and a spatial resolution of 2$^\prime$$^\prime$ per pixel. In addition, we also use the data of soft X-ray flux observed by Geostationary Operational Environmental Satellite (GOES) to identify flare occurrence. The data from Large Angle and Spectrometric Coronagraph (LASCO; Brueckner et al. 1995) C2 on board SOHO (Domingo et al. 1995) are used to identify the coronal mass ejection (CME).

\section{Sunspot rotation and the magnetic field evolution}
\subsection{Sunspot rotation}
Figure 1 shows the whole NOAA AR 08858 observed by TRACE white-light (left panel) and SOHO/MDI magnetogram (right panel). The rotating sunspot is marked by the red box and the black arrows in Fig. 1. The area of the red box and the yellow box is used to calculate the positive and the negative magnetic flux. The rotating sunspot with positive polarity had two umbrae signed by umbra 1 and umbra 2. This active region contains twelve sunspots. The rotating sunspot was the largest one and located in the southeast of the active region.

Figure 2 shows the evolution of the rotating sunspot acquired at white-light, 1600 \AA\, and 171 \AA\ by TRACE. The left column of Fig. 2 shows the white-light images observed by TRACE. We mark the two umbrae as ``U1" and ``U2". From 00:00:22UT to 02:40:36UT on February 9, the rotating sunspot was almost quiet. Later, the two umbrae began to rotate counterclockwise. The detailed motion in the photosphere can be seen from the change of the positions of ``U1" and ``U2". Following the sunspot rotation, the loop-shaped structure first appeared in the chromosphere and then formed an arch-shaped structure. In the corona, the filament was gradually formed in the filament channel. The filament connecting the rotating sunspot was also found to rotate around the center of rotating sunspot counterclockwise.

Figure 3 shows the three images acquired at white light, 1600\AA, 171\AA\ on February 9. The circles in the images contain two umbrae of the rotating sunspot and are used to calculate the rotational angle. The white brackets denote the rotational angles. The arrows denote the features which are used to calculate the rotational angle. We calculated the rotational angle of umbra ``U2" around the center of circle. The front of umbra ``U2" (see the arrow in the left panel of Fig. 3) that moved along the circle is used to evaluate the rotational angle (see the left panel of Fig. 3). From a series of the TRACE images, we can get the coordinates of the center of the circle and the points. We adopt the average values of three repeated measurements of the angles. The measurement uncertainty is about one degree. Moreover, the rotational angles of both the bright loop-shaped structure marked by the arrows in TRACE 1600 \AA\ images and the filament marked by the arrows in TRACE 171 \AA\ images were also calculated. The emitting structure is identified as filament whereas the absorptive dark structure is identified as filament channel in the right panel of Fig. 2. We use the part of the bright loop-shaped structure (see the arrow in the middle panel of Fig. 3) and the filament that connected the umbrae of the rotating sunspot (see the arrow in the right panel of Fig. 3) to calculate the rotational angle. We trace the evolution of the features from a series of TRACE 1600 \AA\ and 171 \AA\ images to determine the positions of the features. Note that the angle is defined as the angle between the line connecting the point where the bright loop-shaped structure is situated on the circle with the center of the circle and the radius of the circle at 0 degree. Because the bright features have a certain width, we adopt the center point of the bright features to do the measurement, which is located on the circle. The radius of the circle is 5 arcseconds. The coordinates can be seen from Fig. 3. It is worth pointing out that the bright loop-shaped structure in the chromosphere and the filament are 3-dimensional and the projection effect has to be taken into account when measuring apparent motion of a feature in general. It is hard to reconstruct the real shape of the loops by the single spacecraft observations. Our observations are based on the evolution of the magnetic loop topology from two-dimensional data.

Figure 4 shows the rotational angle of umbra ``U2" (red line), the bright loop-shaped structure in TRACE 1600 \AA\ (blue line), the filament in 171 \AA\ images (green line) and the evolution of the magnetic flux (negative: dashed line; positive: dotted line). The umbra ``U2" rotated by 195 degrees. The average rotation rate was about 8 degrees per hour for 24 hours. The fastest rotation in the photosphere took place during 14:00UT to 22:01UT on February 9, with a rotation rate of nearly 16 degrees per hour. The bright loop-shaped structure in the chromosphere and the filament in the corona rotated by 142 degrees and 116 degrees. From 15:28UT to 19:00UT, the bright loop-shaped structure in the chromosphere and the filament in the corona rotated by 65 degrees and 85 degrees, with a rotation rate of nearly 19 degrees and 24 degrees per hour. Diamond (green line), Asterisk (blue line) and Plus (red line) respectively denote the rotational angles of the filament in the corona, the bright loop-shaped structure in the chromosphere, and umbra ''U2" in the photosphere. The rotational angle decreased from the photosphere to the corona. It is evidenced that the sunspot rotation transfers the magnetic twist from the sub-surface to the corona.

\subsection{The magnetic field evolution}

The dashed line and the dotted line in Fig. 4 show the evolution of magnetic flux (right axis) calculated from the region marked by the yellow (negative magnetic flux) box and the red (positive magnetic flux) box. From the evolution of the magnetic flux, there was a slow decrease of negative magnetic flux from 11:15UT to 20:47UT on February 9 and then the negative magnetic flux increased a little. For the positive magnetic flux, there was a slow increase from 23:59UT on February 8 to 14:27UT on February 9 and then the positive magnetic flux increased rapidly from 16:03UT to 19:11UT on February 9. Interestingly, the rapid increase of the positive magnetic flux just occurred during the fastest rotation of the rotating sunspot, the bright loop-shaped structure and the filament. At the beginning of the sunspot rotation, the magnetic flux was very stable. In addition, there was no eruption within about five hours from GOES observation before the sunspot rotation. The disturbance from the eruptions in this active region can also be excluded.

\section{The formation and eruption process of the filament}
\subsection{The formation process of the filament}
 From 02:49:36UT on Feb. 9 to 02:49:45UT on Feb. 10, 2000, the two umbrae rotated counterclockwise by 195 degrees. The middle column of Fig. 2 shows the 1600 \AA\ images observed by TRACE. There was a small bright loop-shaped structure marked by the white arrows at 02:40:32UT in TRACE 1600 \AA\ images. The loop-shaped structure was followed by the sunspot rotation and rotated counterclockwise around the center of the rotating sunspot. From 16:19:54UT to 20:35:48UT on Feb. 9, the loop-shaped structure formed an arch shape. Finally, it disappeared after the flare. The right column of Fig. 2 shows the 171 \AA\ images observed by TRACE. The dotted lines in the first two images of the right column denote the EUV filament channel. The red dotted lines indicate the filament. Until 12:17:46UT on Feb. 9, a curve loop-shaped filament marked by the white arrows and outlined by the red dotted line appeared. The filament also rotated counterclockwise around the center of the sunspot. The filament was formed as a dark structure initially, and then part of it was brightened. This brightened part connecting the rotating sunspot was identified and measured. The change of the filament can be seen from the positions marked by the white arrows in the right column. Following the sunspot rotation, the part of the filament that connected the umbra of the rotating sunspot met the left part of the filament channel (see the position marked by the black arrow in the right panel of Fig. 2), then the rotational motion stopped and the filament finally erupted. The field of view of the left and the middle column images is 50$^\prime$$^\prime$ $\times$ 50$^\prime$$^\prime$. In order to show the formation process of the active-region filament, the field of view of the right column images is adjusted to 150$^\prime$$^\prime$ $\times$ 150$^\prime$$^\prime$. The detailed formation process of the filament can be seen from the movie (filamentformation.mpg) linked to Fig. 2.

\subsection{The first failed eruption of the filament}

Figure 5 shows a sequence of 171 \AA\ images during the first failed filament eruption on February 10, 2000. The dashed line in Fig. 5a indicates the filament channel and the white line in Fig. 5a denotes the position of the time slice of Fig. 6. From a sequence of TRACE 171\AA\ images, one can see that the filament gradually rose from the EUV filament channel after the sunspot underwent tens of hours of rotation motion. The filament was marked by the dotted lines in Figs. 5b and 5c. Moreover, the filament exhibited a swirling shape. The white arrows in Fig. 5b and 5c point to the hot material of the filament. Before the filament eruption, the filament loops carrying the hot material can be seen to be moving counterclockwise. After about ninety seconds, the hot plasma moved to the position signed by the white arrow in Fig. 5c. At 01:14:38UT, the filament rose rapidly and formed a fan-shaped structure. The two dotted lines outline the outer and the inner boundary of the filament in Fig. 5d. Note that the filament is composed by many bright loops. The white arrow and the black arrow in Figs. 5d and 5e denote the lower and the upper part of the filament. During the rise of the filament, we observe apparent counterclockwise motion of hot and cool materials along the filament loops. The black arrows in Figs. 5f-5i denote the change of the position of the cool material. The white arrows denote the hot material which gradually fell into the umbrae of the sunspot. It is worth pointing out that the movement of the cool and the hot material is not the true movement of the material. In fact, the movement of the filament loops carried the cool and the hot material. At 01:26:48UT on February 10, 2000, the loops of the whole filament were contracted and later were seen to spiral into the sunspot umbrae. The dotted lines in Figs. 5e-5l outline the outer boundary of the filament. From 01:14:38UT to 01:26:48UT on February 10, the filament loops gradually contracted (see the dotted lines in Fig. 5). The detailed process of the first filament eruption can be seen from the movie (firsteruption.mpg) linked to Fig. 5.

Figure 6 shows the time slice at the position marked by the white line in Fig. 5a. The bright structure shows the trajectory of the filament. The two dotted lines denote the lower and the upper boundary of the filament. From the evolution of the filament intensity, one can see the filament first expanded outward and then fell down.

\subsection{The second successful eruption of the  filament}
After the first failed eruption, the filament gradually became active. At 01:36:16UT on February 10, a small part of the filament began to erupt. From the observation of TRACE 1600 \AA\ (see Fig. 9), the two flare ribbons began to form at 01:40UT as signature of magnetic reconnection upon the filament eruption. Figure 7 shows a sequence of 171 \AA\ images from 01:38:24UT to 02:39:05UT on February 10, 2000. At 01:38:24UT on February 10, the right part of the filament became active again. The white arrow points to the same loop of the filament in Figs. 7a-7c. At 01:39:59UT, another bright loop marked by the black arrow in Fig. 7b appeared. Subsequently, the bright loop marked by the black arrow first disappeared and then the other bright loop marked by the white arrow vanished. The disappearance of the features may be temperature effect. After the eruption of the two bright loops, the bright material of the filament was found to flow from right to left. The white arrows in Figs. 7d-7g indicate the positions of the hot plasma from 01:43:39UT to 01:48:23UT on February 10. The filament loops carrying the hot plasma gradually moved along the loop from west to east. At 01:48:56UT, another part of the filament enclosing the rotating sunspot also erupted. Next, the filament exhibited clearly an inverse S-shaped structure marked by the dotted lines in Figs. 7i and 7j. There was a data gap from 02:08:05UT to 02:35:22UT on February 10. However, comparing the change of the magnetic structure, it is easy to find that the inverse S-shaped filament disappeared. The post-flare loops marked by the white arrows in Figs. 7k and 7l can be seen clearly. The detailed process of the second filament eruption can be seen from the movie (seconderuption.mpg) linked to Fig. 7.

\subsection{The associated flare and CME}

Figure 8 shows the evolution of GOES soft X-ray emission for the C7.3 flare on February 10, 2000. The C7.3 flare started at 01:40UT, peaked at 02:08UT, and ended at 02:39UT. Figure 9 shows the evolution of the flare ribbons from 01:41:39UT to 02:35:28UT on February 10. The two white arrows in Fig. 9a and 9b indicate the two flare ribbons. The two flare ribbons gradually became brightening. The left flare ribbon along the dotted lines in Figs. 9c-9g expanded toward the southwest of the following sunspot. The flare ribbons swept across the umbra of the following sunspot while this did not happen to the leading rotating sunspot. Finally, the flare ribbon swept completely the following sunspot. Li \& Zhang (2009) suggested that the emergence, the rotation, and the shear motion of the following sunspot and leading sunspot caused flare ribbons to sweep across sunspots completely. In this event, the flare ribbon did not sweep the rotating sunspot unlike those examples that Li \& Zhang (2009) investigated. After the inverse S-shaped filament erupted, the SOHO/LASCO observed a halo CME.

\section{Conclusion and Discussion}
We investigate the relationship between the sunspot rotation and the formation and eruption of an active-region filament associated with a C7.3 flare and a
halo CME in NOAA AR 08858 on Feb 10, 2000 using the GOES12 soft X-ray flux, TRACE WL, 1600 \AA\
and 171 \AA\ images, SOHO/MDI 96-min magnetograms, and SOHO/LASCO C2 images. We find that the formation of the active-region filament in EUV filament channel was followed by sunspot rotation. The sunspot rotated counter-clockwise and the active-region filament exhibited an inverse S-shaped structure. The filament experienced two eruptions. In the first eruption, a part of the filament rose and much of the material warmed up (becoming bright). The filament loops carrying the material were seen to spiral into the sunspot counterclockwise in the middle as it fell back towards the solar surface. In the second eruption, the inverse S-shaped filament fully erupted and produced a C-class flare and a halo CME. Before the second eruption, the filament loops carrying the hot material moved clockwise along the magnetic loop.

This event is a clear case of the formation of the sigmoidal active-region filament caused by sunspot rotation. According to the sunspot rotational direction (counterclockwise) and the shape of the filament (the inverse S-shaped filament), we can determine that the sunspot had negative helicity. The inverse S-shaped filament followed the hemisphere helicity rule. From the topology evolution of the magnetic loops in the corona, one can see that the sunspot rotation resulted in the upper magnetic field rotation and made the magnetic fields trend to non-potential field. It is evidenced that sunspot rotation is a means of magnetic energy storage. The energy was released via flares and CME later.

During the observation, no obvious magnetic flux emergence was found before the sunspot rotation. But there was a slow increase of the positive magnetic flux from 23:59UT on February 8 to 14:27UT on February 9 and then the positive magnetic flux increased rapidly from 16:03UT to 19:11UT on February 9. It is interesting that the rapid increase of the positive magnetic flux just occurred during the fastest rotation of the rotating sunspot, the bright loop-shaped structure and the filament. The observation provides evidence that the sunspot rotation could be regarded as a result of the transfer of additional magnetic twist from the sub-surface to the corona. The investigation of Zhao \& Kosovichev (2003) also evidenced that there was a strong subsurface vortical flow below a rotating sunspot. Magara \& Longcope (2001) and Fan (2009) presented a simulation on the emergence of a twisted flux tube into the solar atmosphere. During the emergence, the opposite polarity regions separated and rotated toward a more axial orientation. Fan (2009) concluded that the rotation in the two polarities is a result of propagation of nonlinear torsional Alfven waves along the flux tube, which transports significant twist from the tube's interior portion to its expanded coronal portion. In some events, the sunspot rotation was obviously accompanied with the emergence flux and polarity separation (Zhang, Li \& Song 2007; Jiang et al. 2011). However, in this event, no these characteristics were found. We assume that the sunspot rotation can originate in two ways. It needs more observations to confirm these results.

When the total twist of the field exceeds a little over one turn, or 2.5$\pi$ (Hood 1991; $Vr\check{s}nak$, $Ru\check{z}djak$ \& Rompolt 1991; Rust et al. 1994; T$\ddot{o}$r$\ddot{o}$k, \& Kliem 2003), the flux rope becomes unstable. Leamon et al. (2003) measured the total twist of 191 X-ray sigmoids and found that most of the sigmoids have a total twist less than one turn. In this event, the sunspot rotated by 195 degrees and the twist was less than the critical value obtained by former authors. However, the filament also erupted finally. We assume that the eruption of the flux rope is relative to not only the twist caused by sunspot rotation but also self-twist before sunspot rotation.

\acknowledgments
The authors thank the referee for very constructive comments and suggestions. The authors thank the TRACE, SOHO, and GOES consortia for their data. SOHO is a project of international cooperation between ESA and NASA. This work is sponsored by National Science
Foundation of China (NSFC) under the grant numbers 10903027, 11078005, 10943002,
Yunnan Science Foundation of China under number 2009CD120, China's 973 project under the
grant number G2011CB811400.

\begin{figure*}
\centering
   \includegraphics[width=7cm]{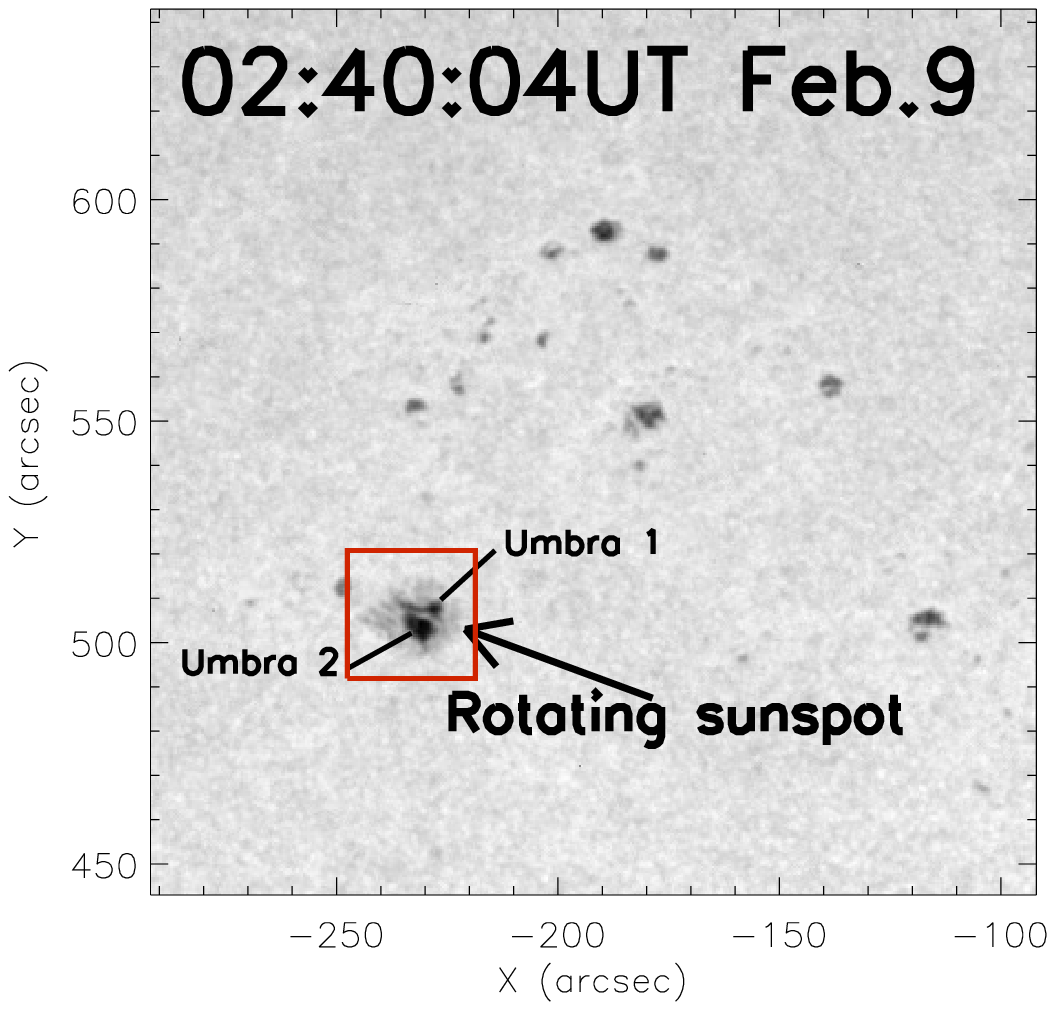}%
   \includegraphics[width=7cm]{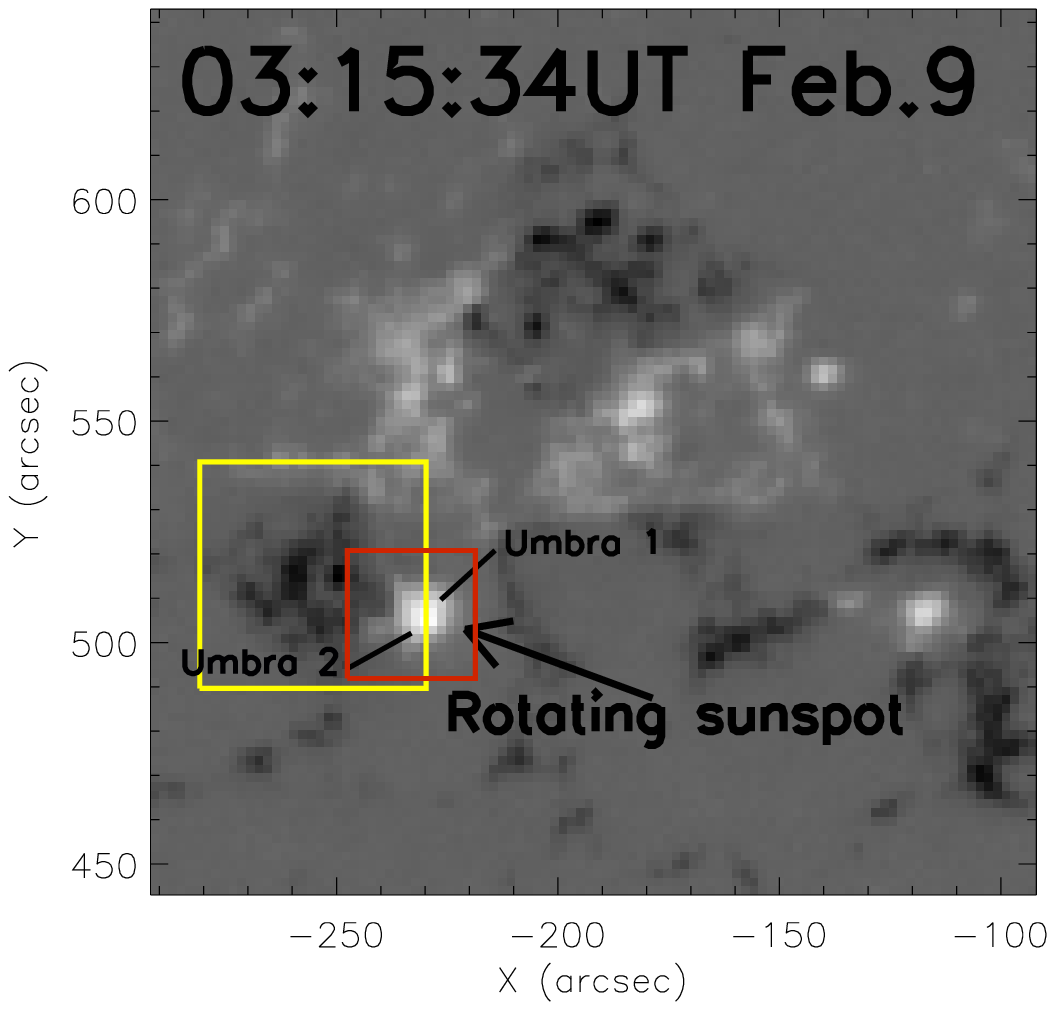}\\
\caption{NOAA AR 08858 observed by TRACE acquired at white-light (left panel) and SOHO/MDI magnetogram (right panel). The rotating sunspot is marked
by the red boxes and the black arrows. The rotating sunpot with positive polarity has two umbrae signed
by umbra 1 and umbra 2. The area of the red box and the yellow box is used to calculate the positive and the negative magnetic flux.}%% no full stop at the end
\end{figure*}

\begin{figure*}
\centering
   \includegraphics[width=6cm]{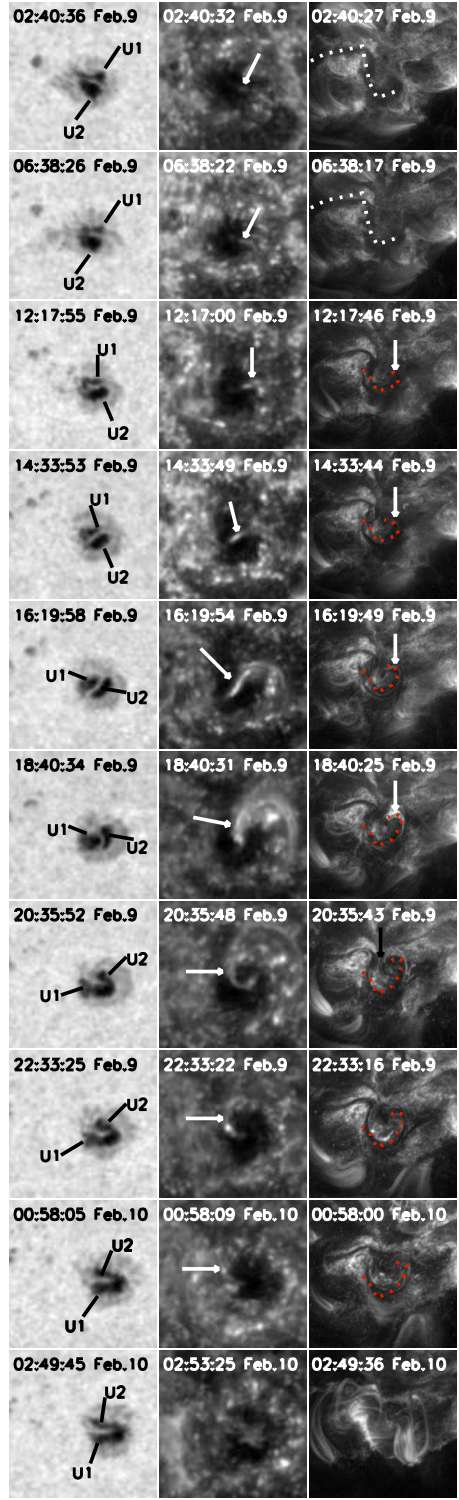}
\caption{The evolution of the rotating sunspot acquired at white-light, 1600 \AA\
and 171 \AA\ observed by TRACE. ``U1" and ``U2" denote the two umbrae of the rotating sunspot. The arrows in the middle column and the right column indicate the bright loop-shaped structure in the chromosphere and the active-region filament in the corona. The white dotted lines indicate the filament channel. The red dotted lines indicate the filament.}%% no full stop at the end
\end{figure*}

\begin{figure*}
\centering
   \includegraphics[width=5cm]{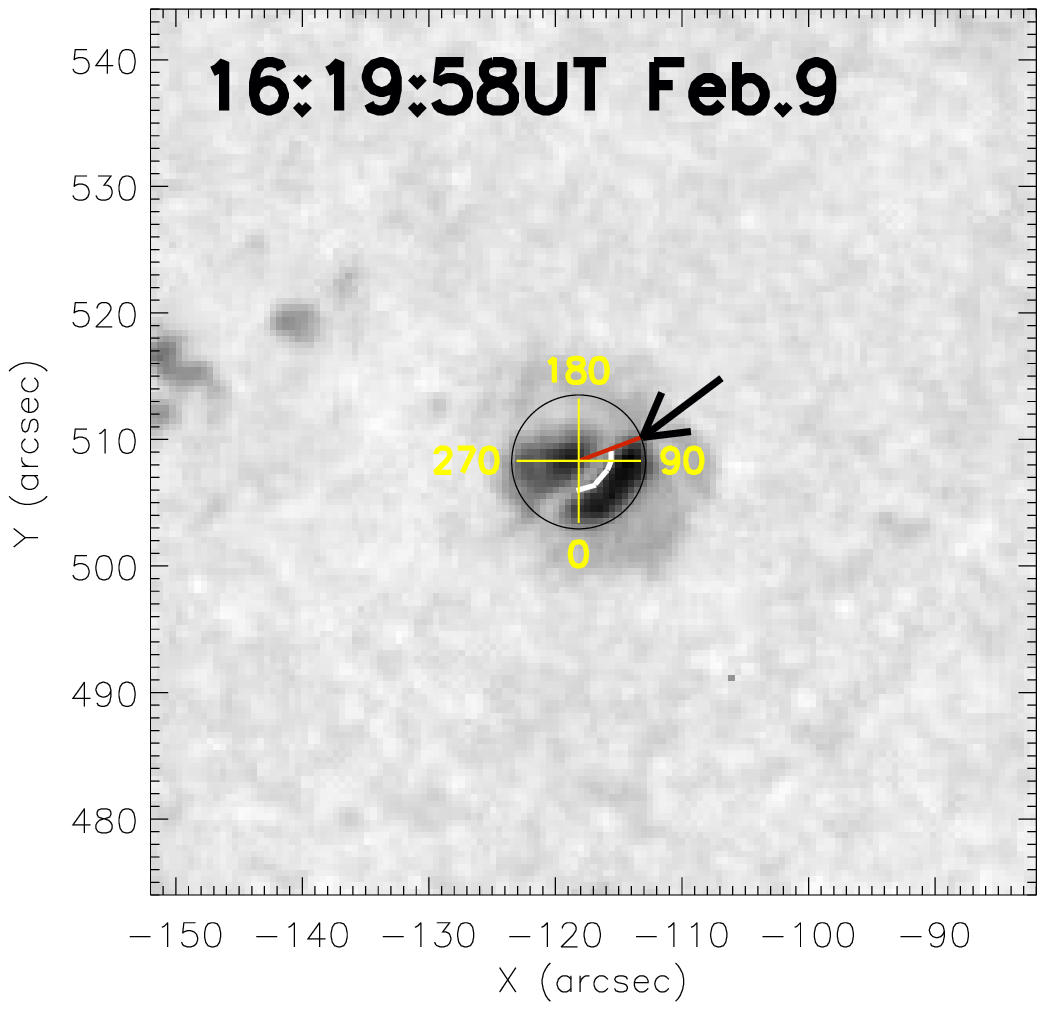}%
   \includegraphics[width=5cm]{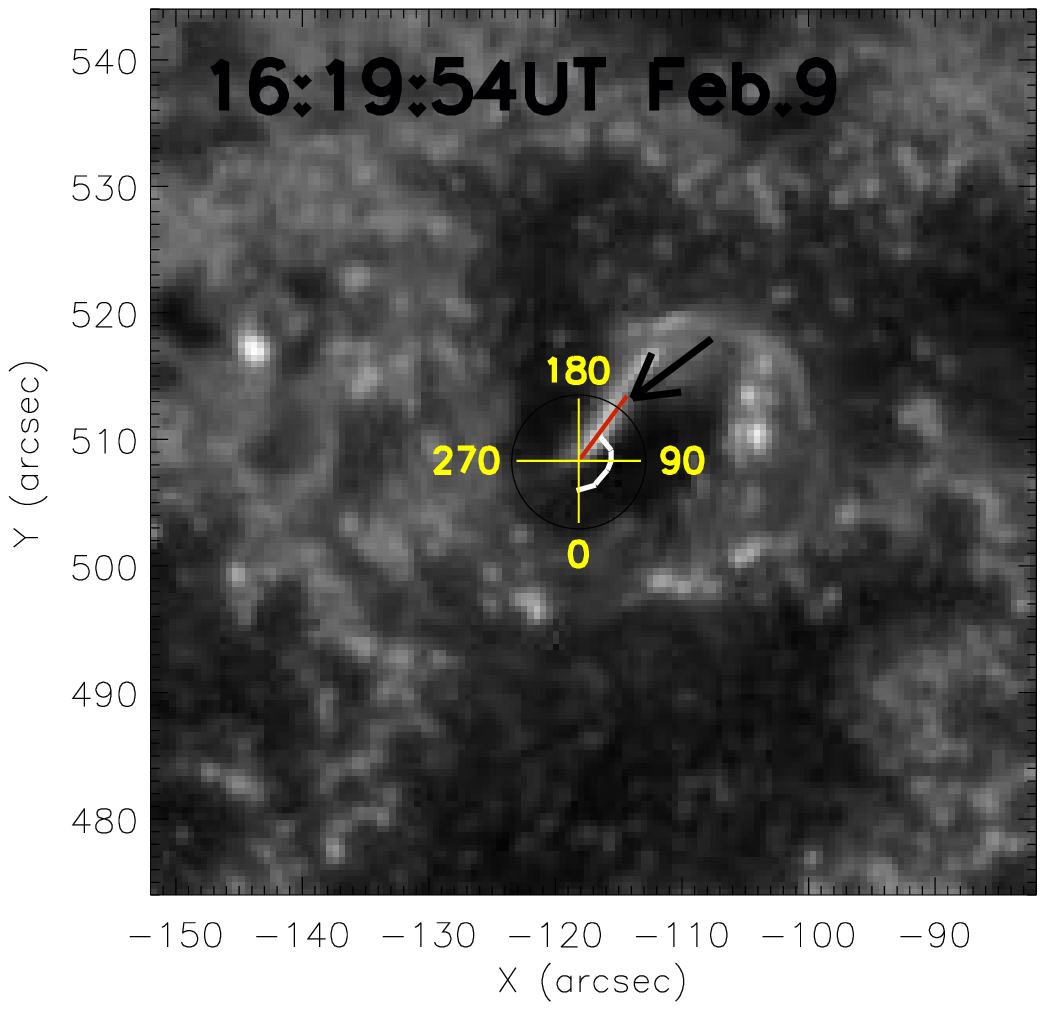}%
    \includegraphics[width=5cm]{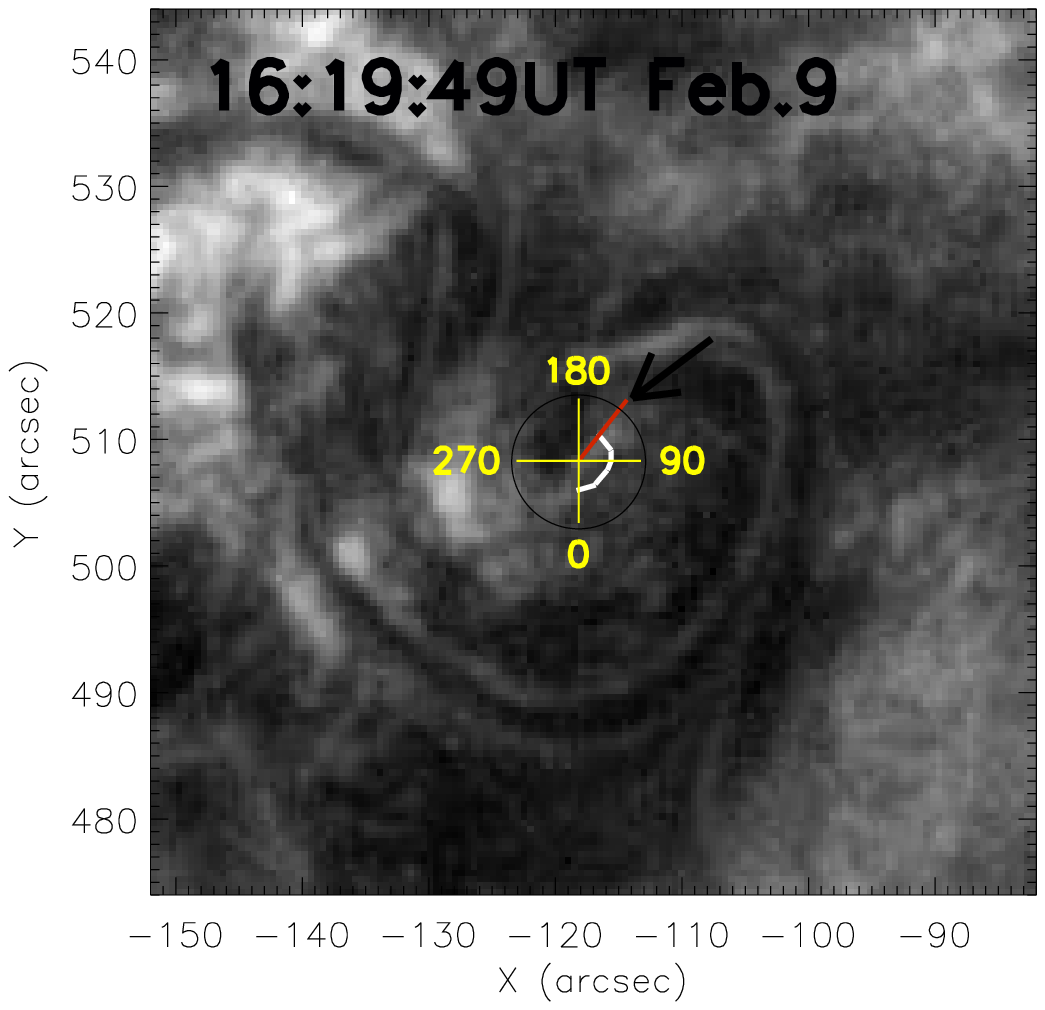}\\
\caption{The three images acquired at white light, 1600 \AA\ and 171 \AA\ (from left to right) on 2000 Feburary 9.
The circles in the images contain the umbrae of the rotating sunspot and are used to
calculate the rotational angles. The bright brackets denote the rotational angles. The arrows denote the features which are used to calculate the rotational angle.}%% no full stop at the end
\end{figure*}

\begin{figure*}
\centering
   \includegraphics[width=10cm]{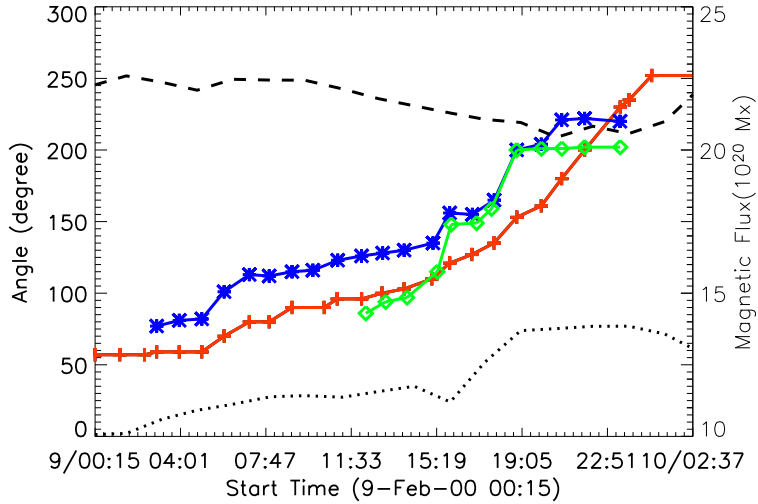}
\caption{The rotational angles measured from umbra ``U2", the bright loop-shaped structure in TRACE 1600 \AA\ image and the filament in 171 \AA\ image. The red line, the blue line and the green line indicate the rotational angle of umbra ``U2" , the bright loop-shaped structure in TRACE 1600 \AA\ image and the filament in TRACE 171 \AA\ image. We adopt the average values of three repeated measurements of the angles. The evolution of the negative (dashed line) and the positive (dotted line) magnetic
flux calculated from the two regions are marked by the yellow and red boxes.}%% no full stop at the end
\end{figure*}

\begin{figure*}
\centering
   \includegraphics[width=4cm]{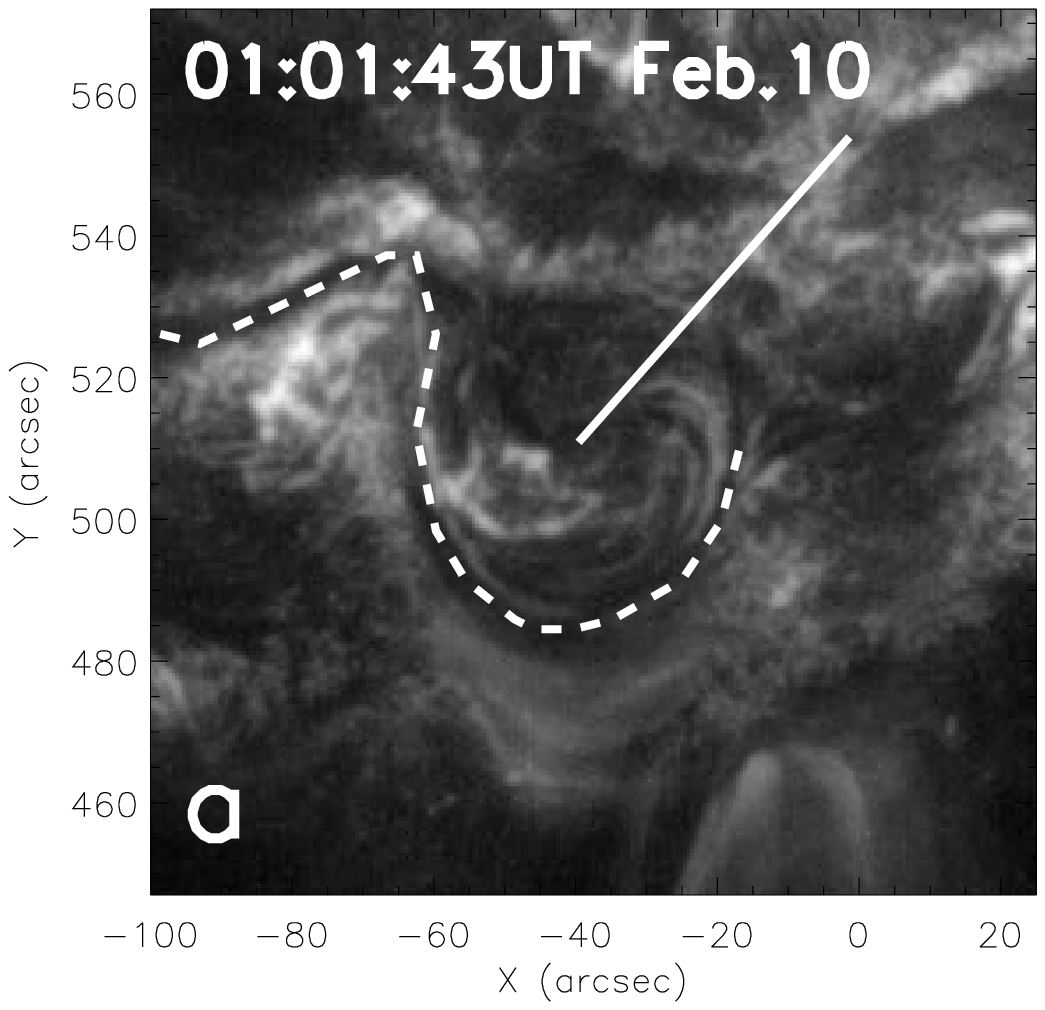}%
   \includegraphics[width=4cm]{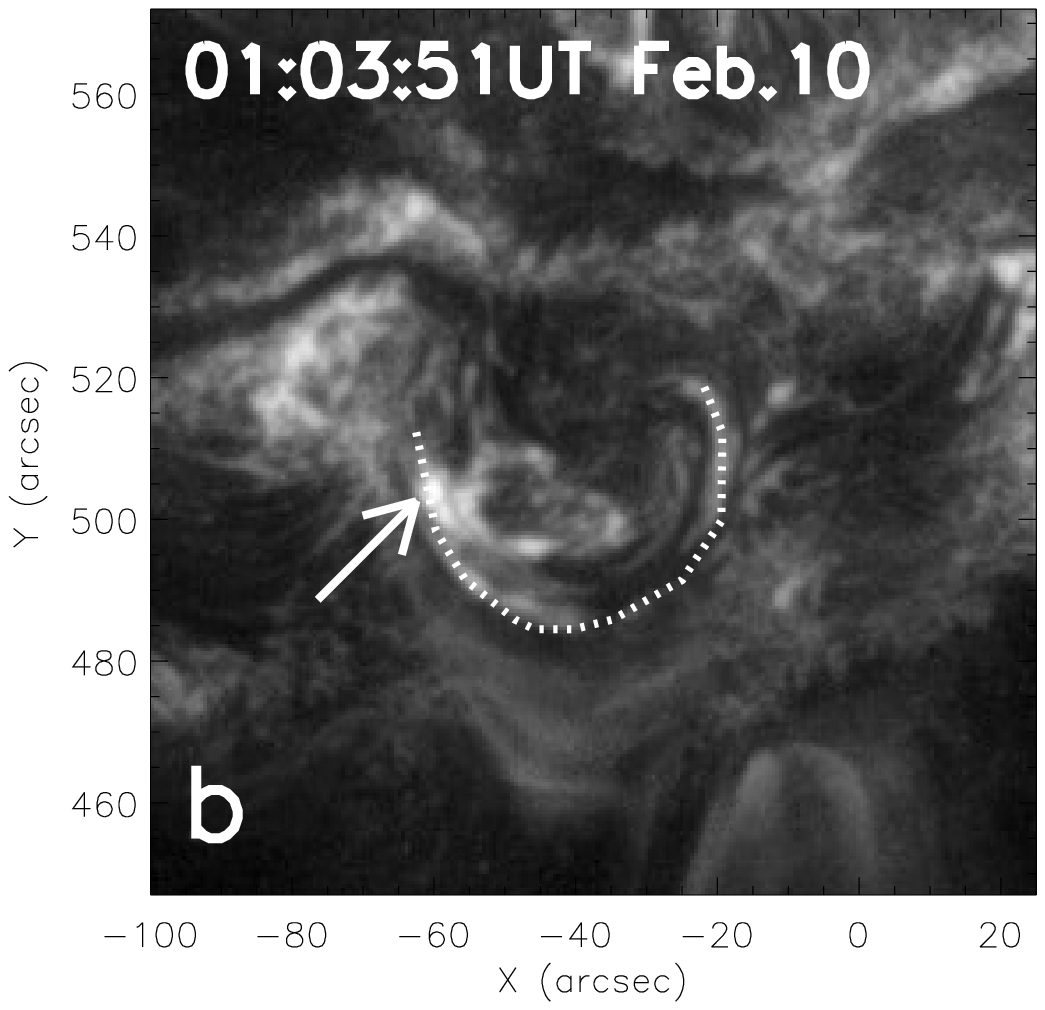}%
    \includegraphics[width=4cm]{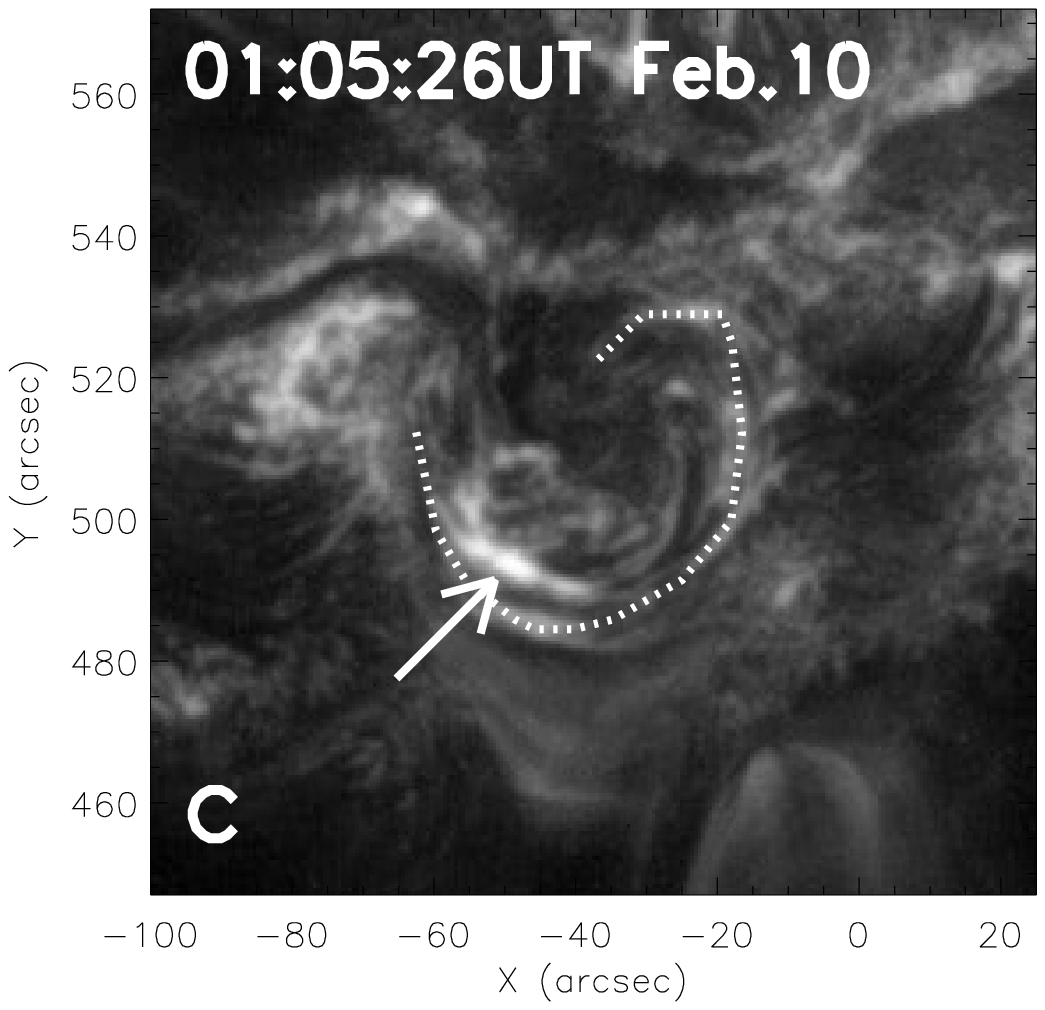}%
    \includegraphics[width=4cm]{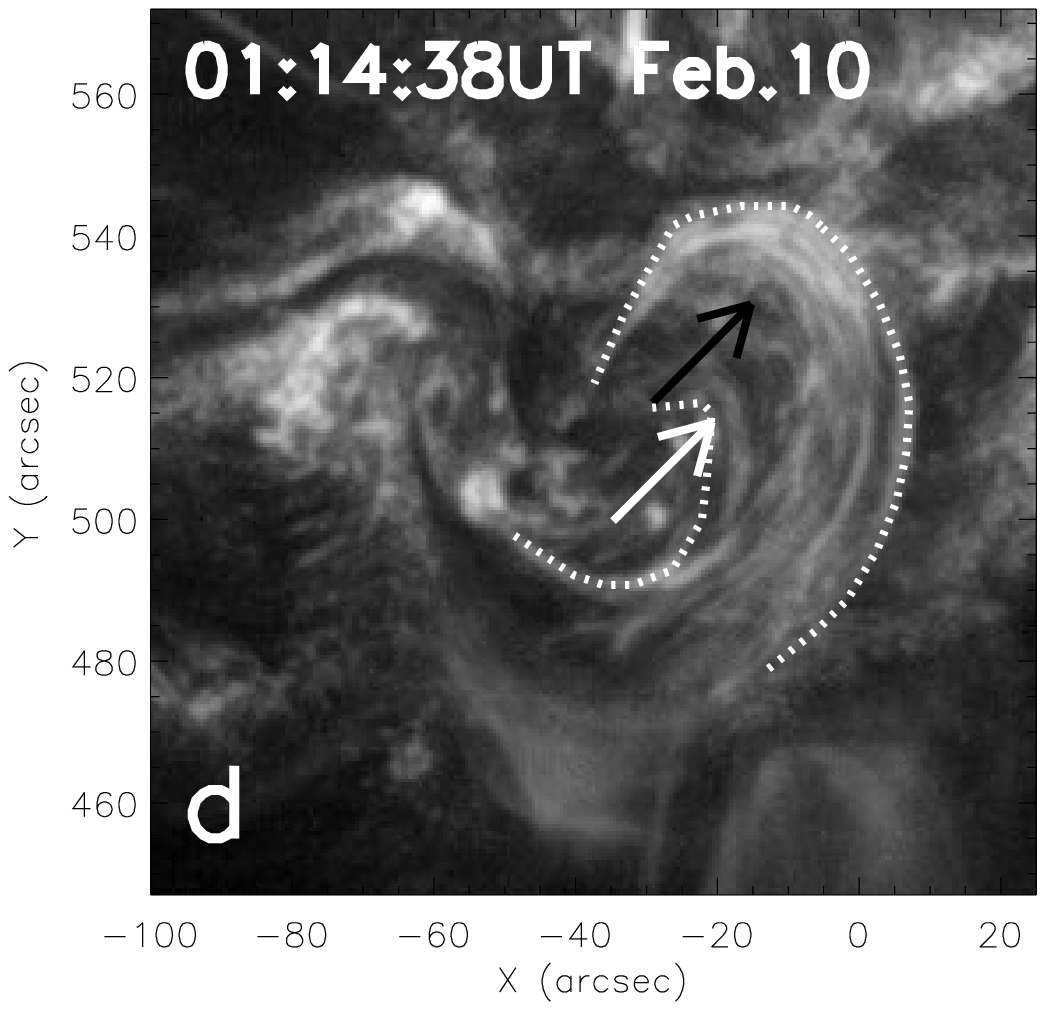}\\
    \includegraphics[width=4cm]{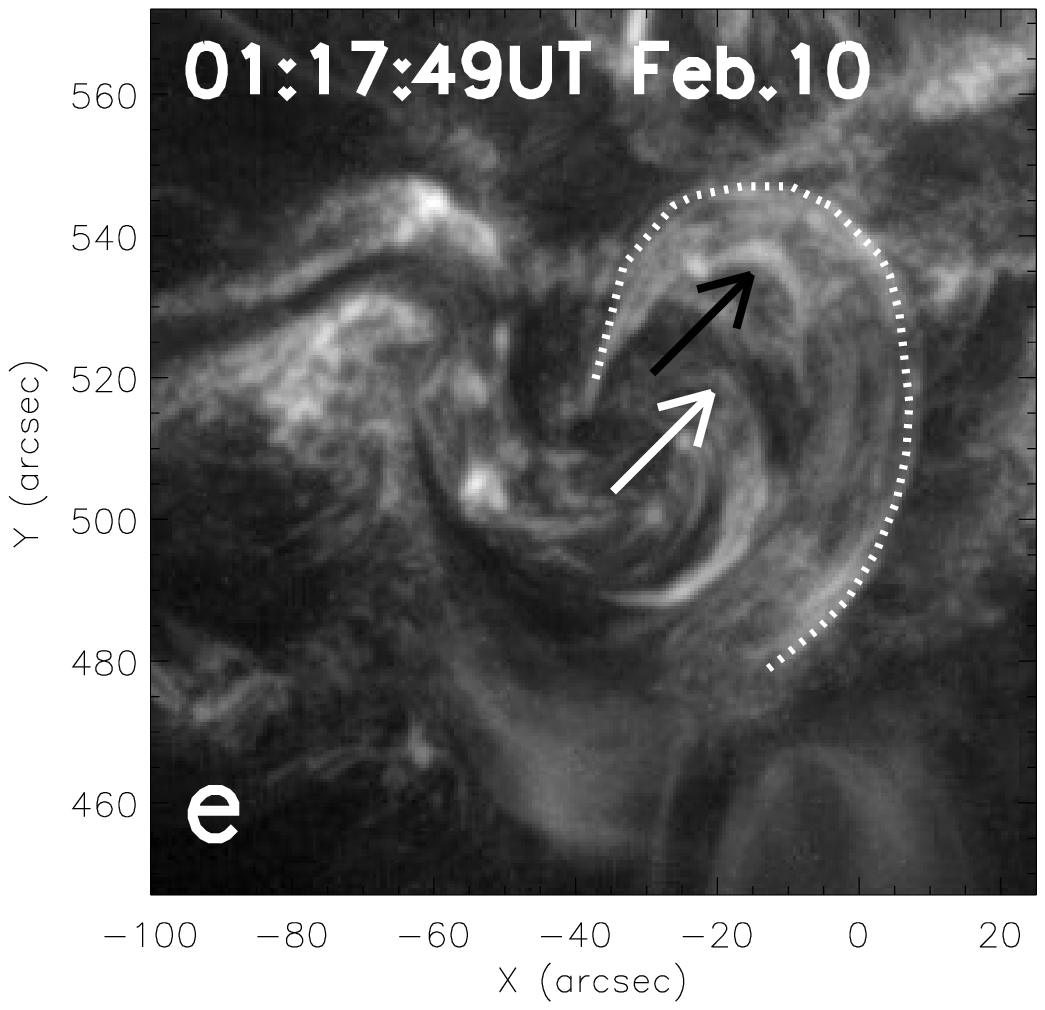}%
   \includegraphics[width=4cm]{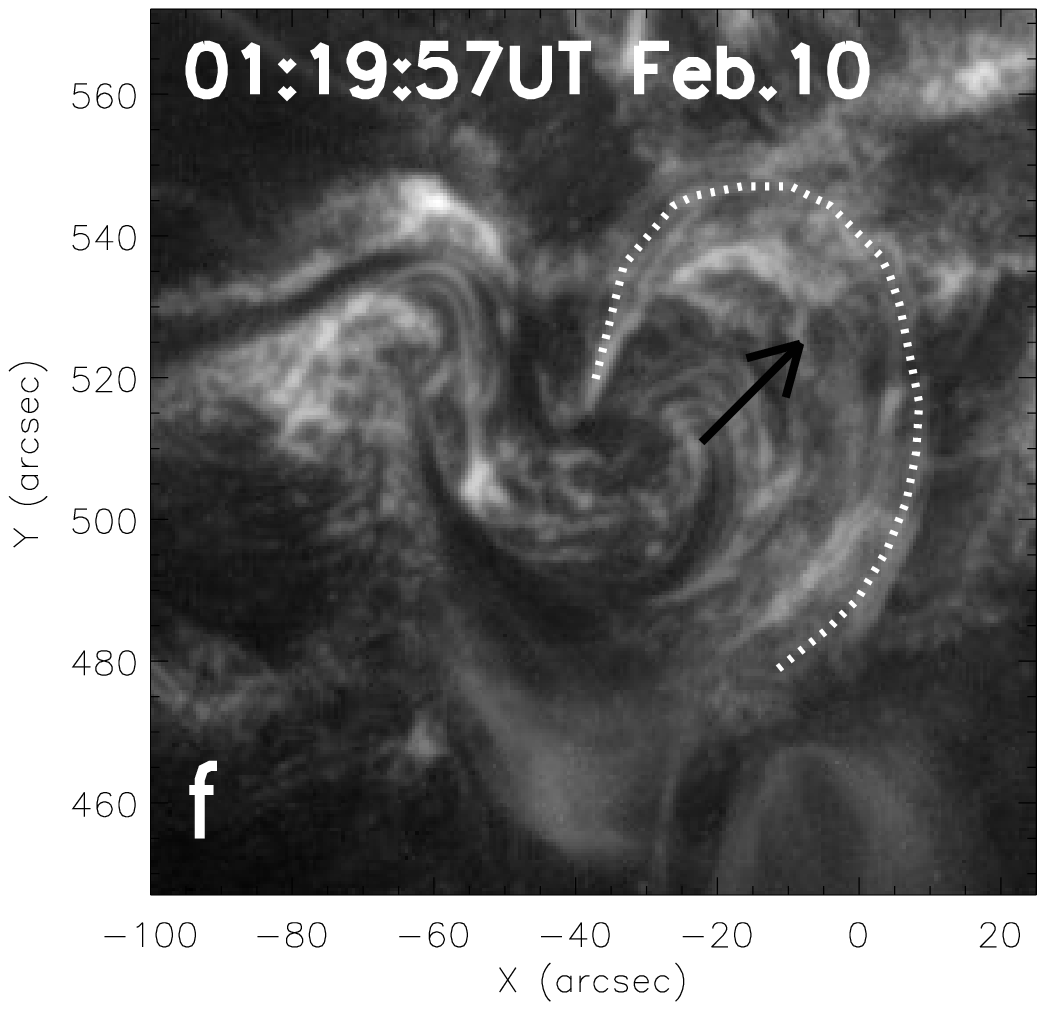}%
    \includegraphics[width=4cm]{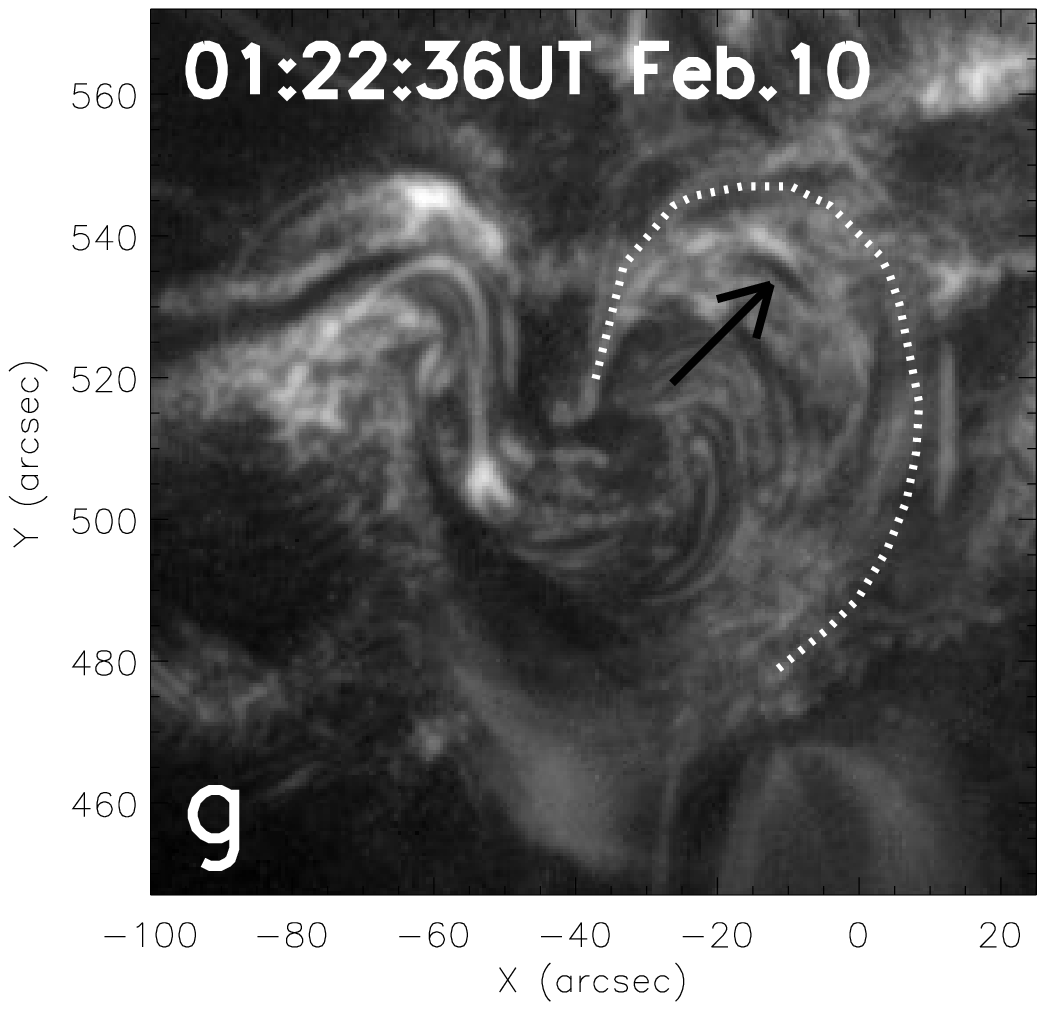}%
    \includegraphics[width=4cm]{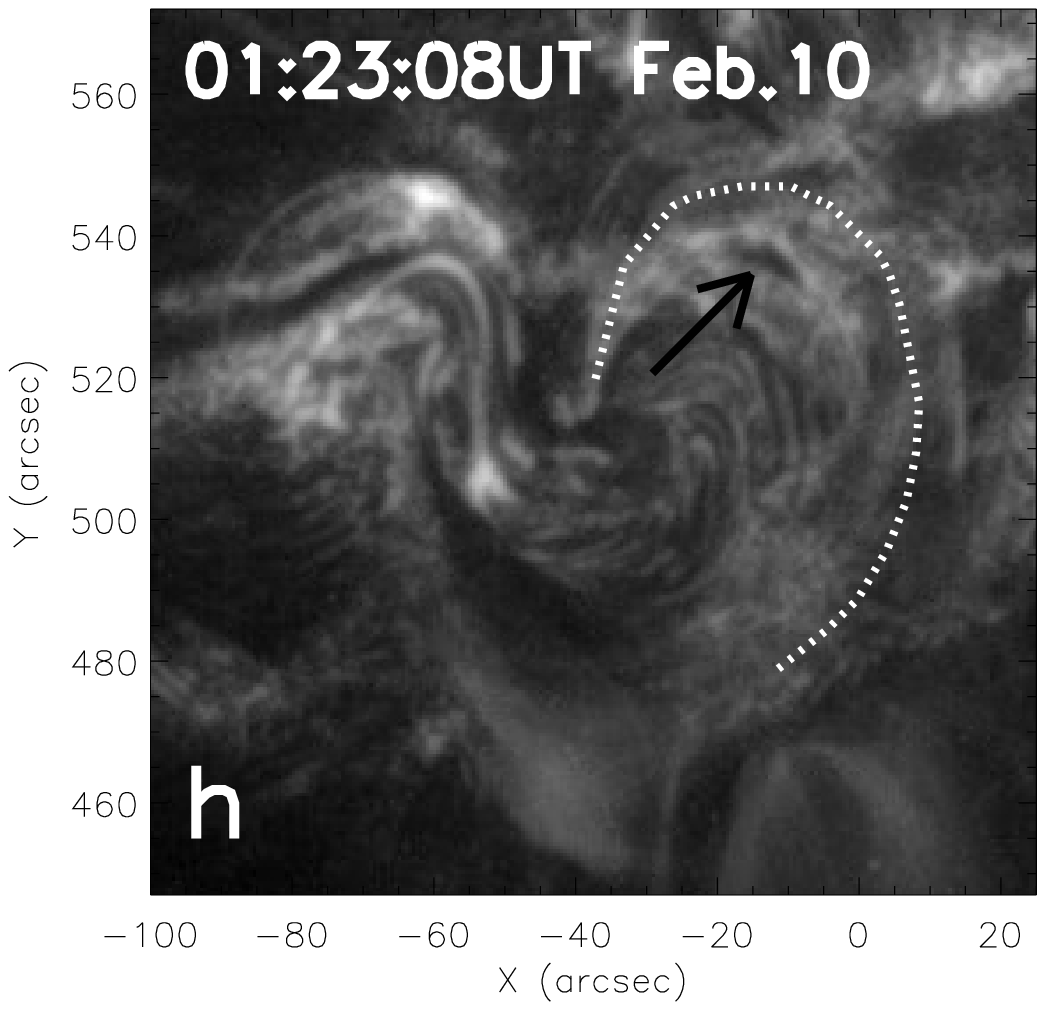}\\
    \includegraphics[width=4cm]{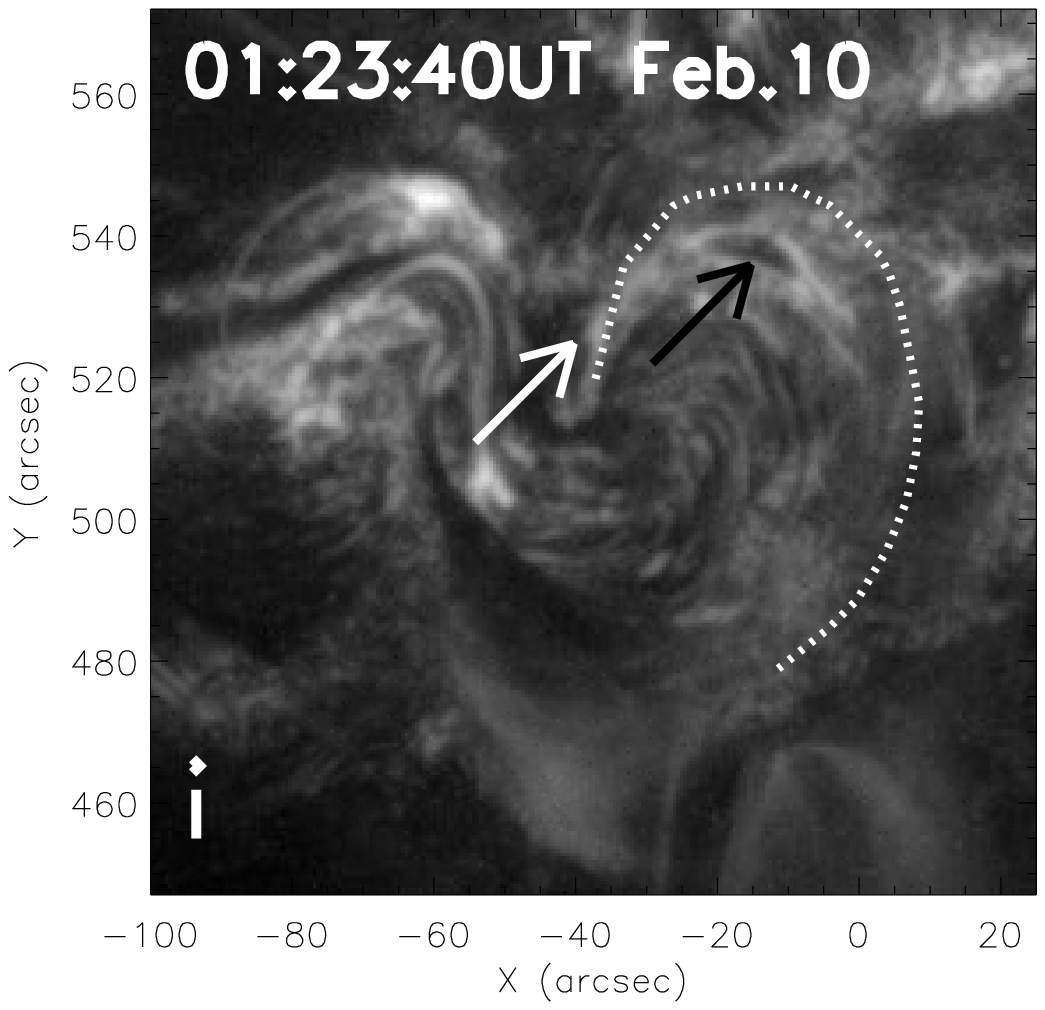}%
   \includegraphics[width=4cm]{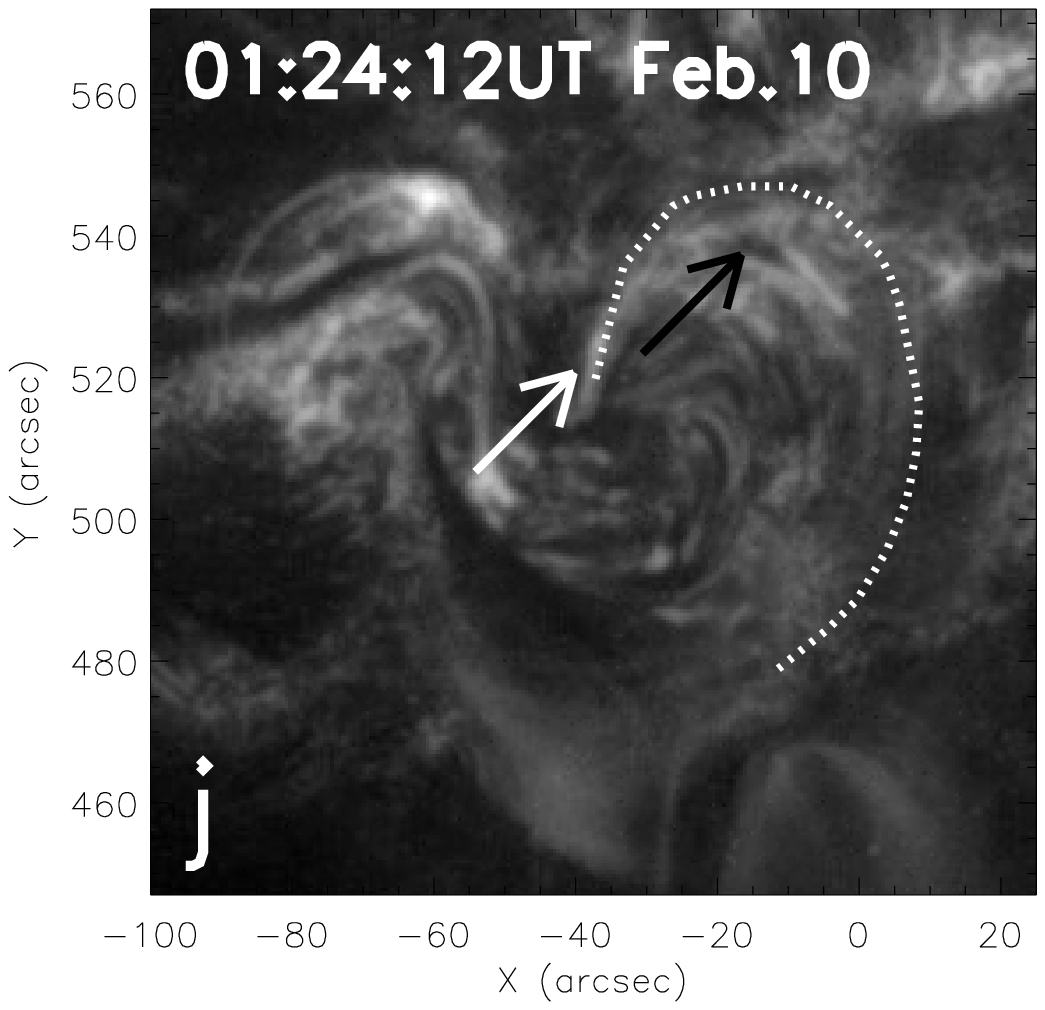}%
    \includegraphics[width=4cm]{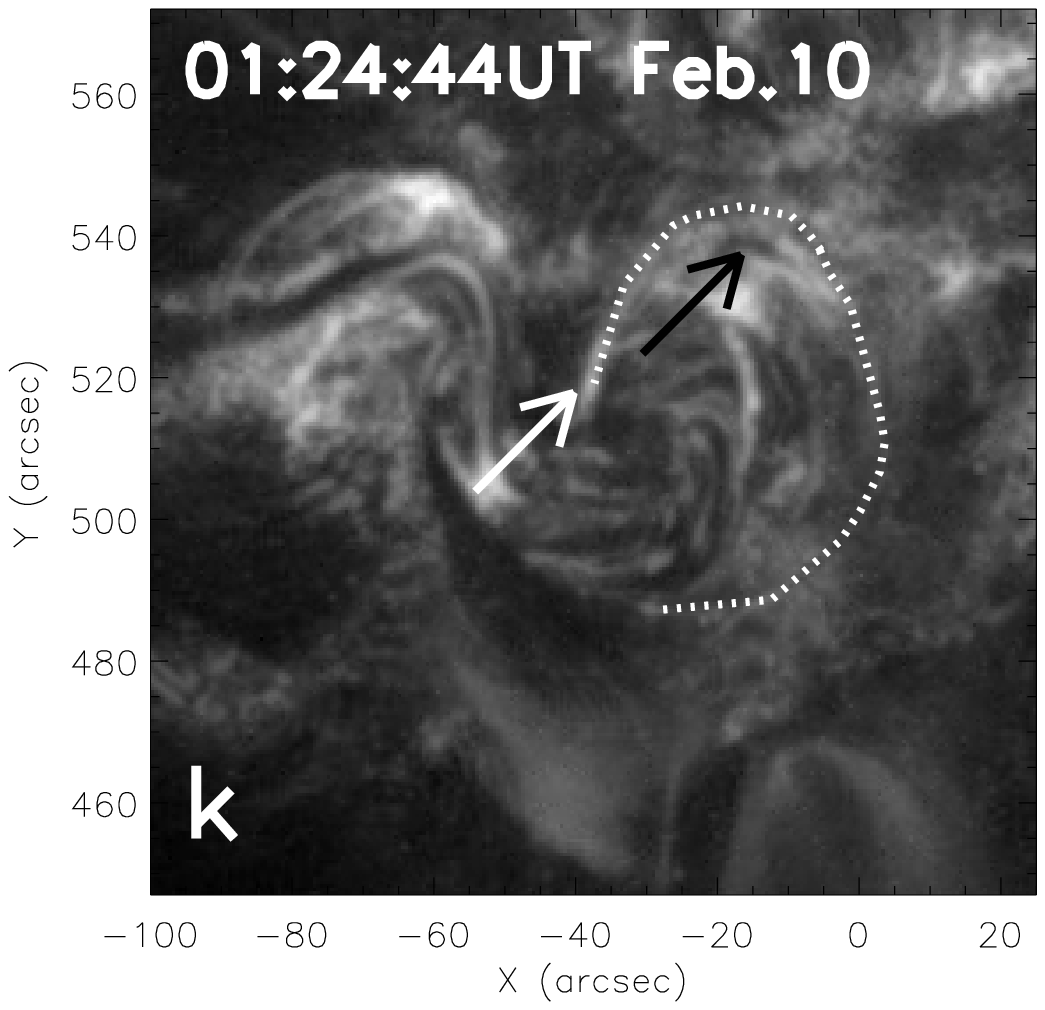}%
    \includegraphics[width=4cm]{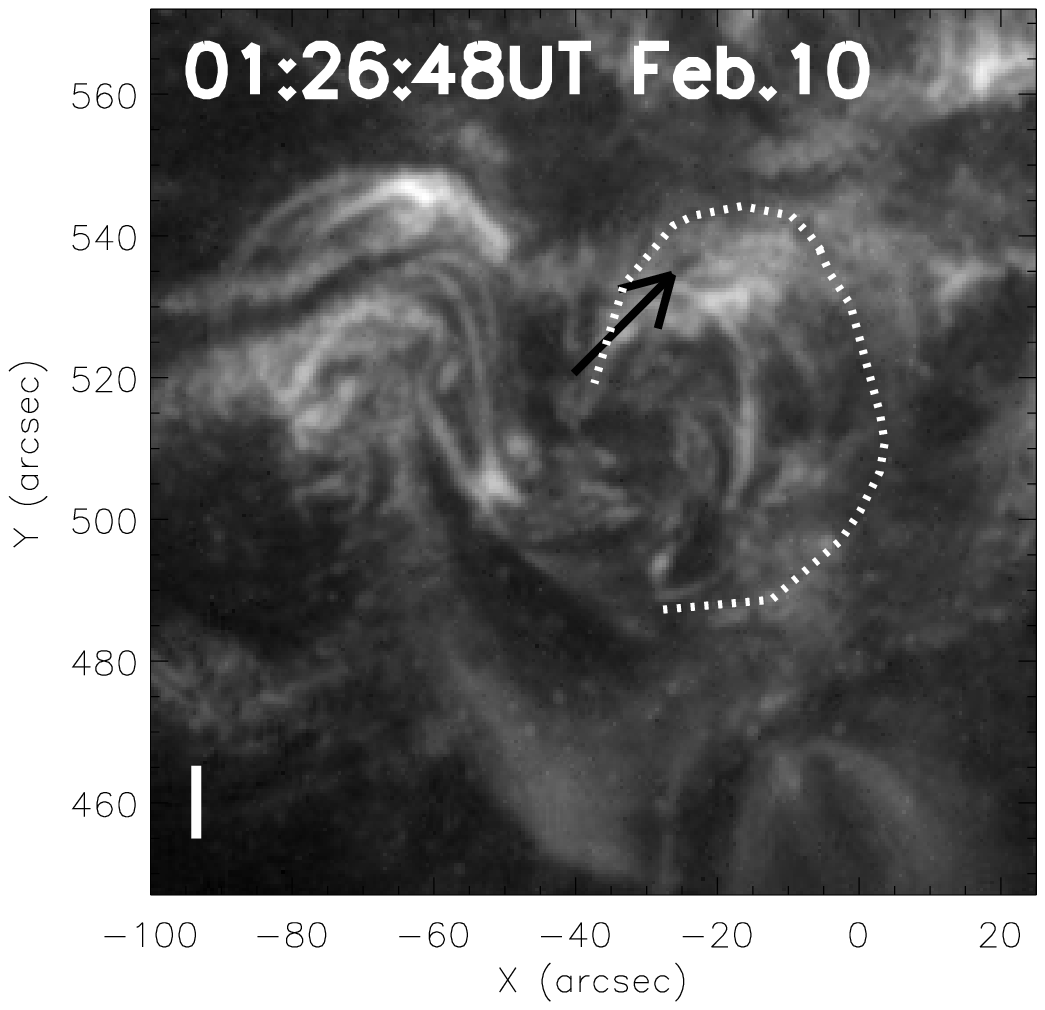}\\
\caption{A sequence of 171 \AA\ images showing the process of the first failed filament eruption.
The dashed line in Fig. 5a indicates the filament channel. The white line denotes the position
of the time slice of Fig. 6. The dotted lines outline the filament in Figs. 5b and 5c. The other dotted lines and the arrows are described in the text.}%% no full stop at the end
\end{figure*}

\begin{figure*}
\centering
   \includegraphics[width=8cm]{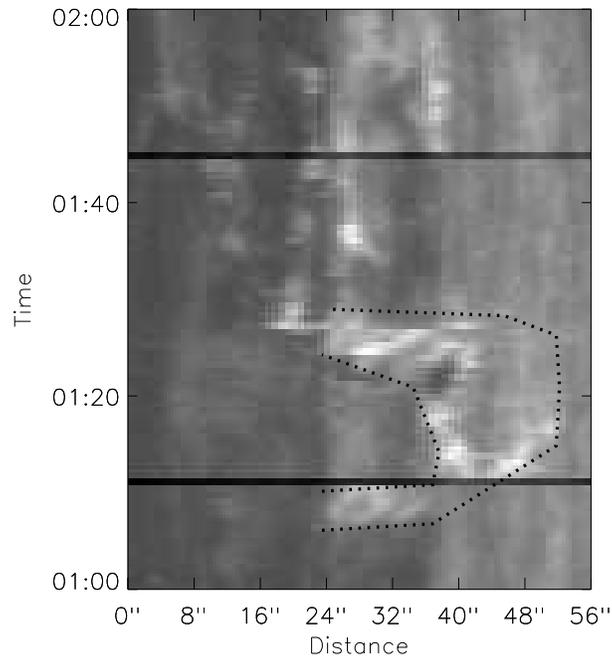}
\caption{The time slice at the position marked by the white line in Fig. 5a. The two dotted lines
denote the lower and the upper boundary of the filament.}%% no full stop at the end
\end{figure*}

\begin{figure*}
\centering
   \includegraphics[width=4cm]{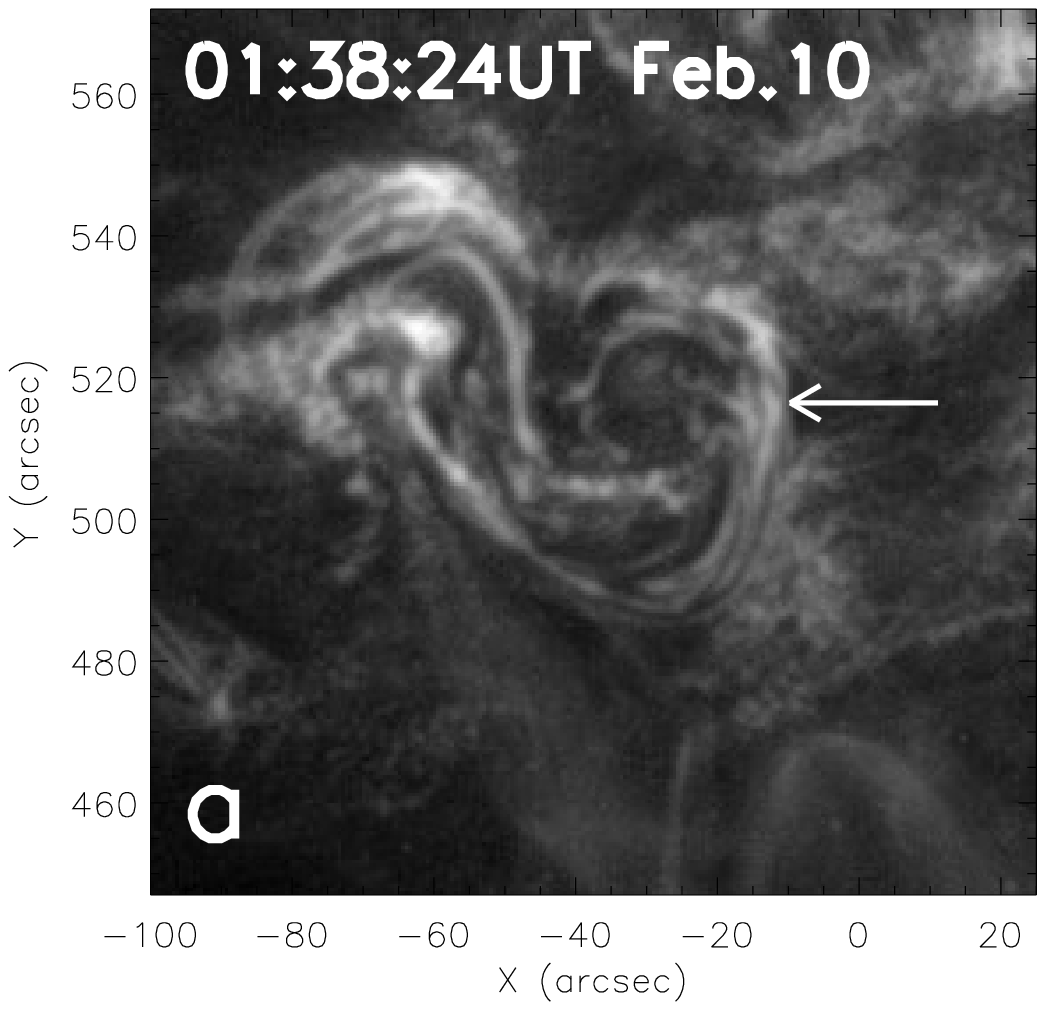}%
   \includegraphics[width=4cm]{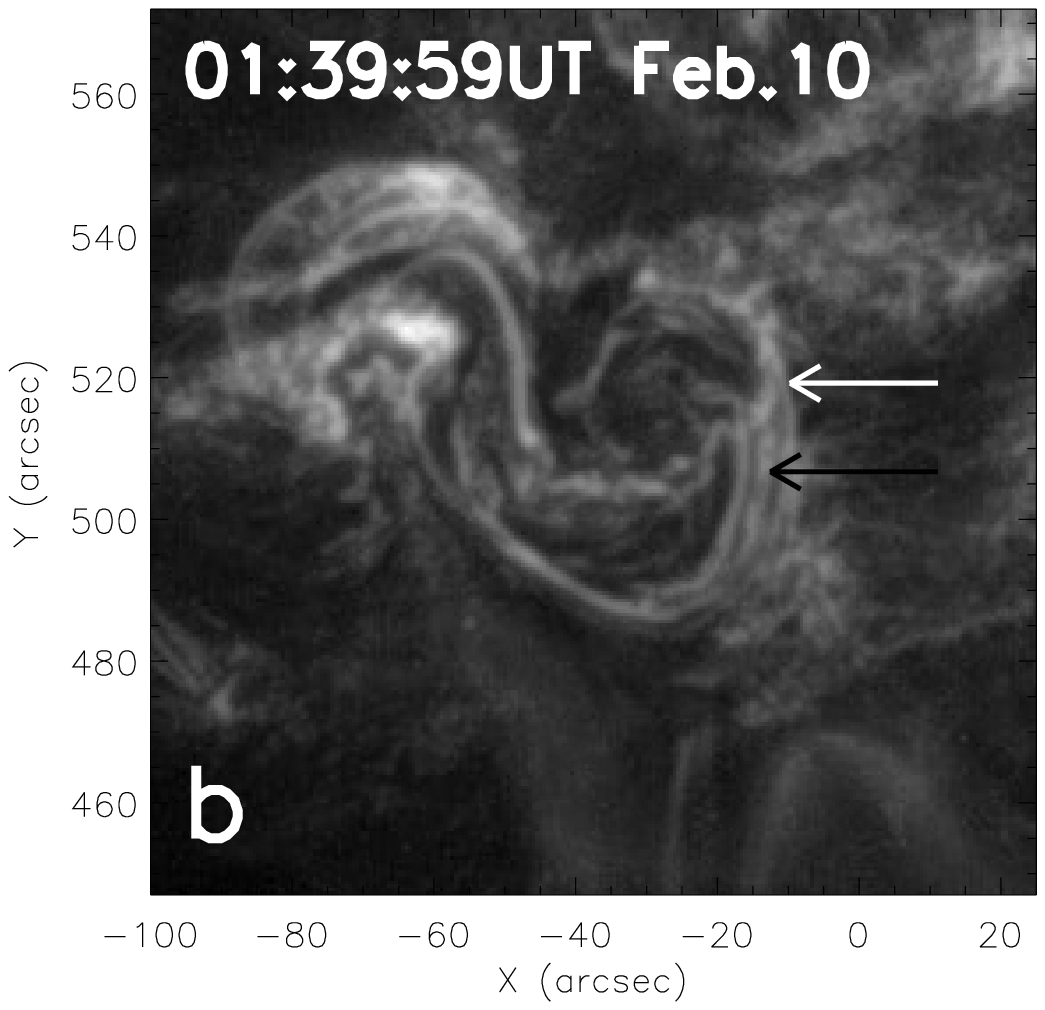}%
    \includegraphics[width=4cm]{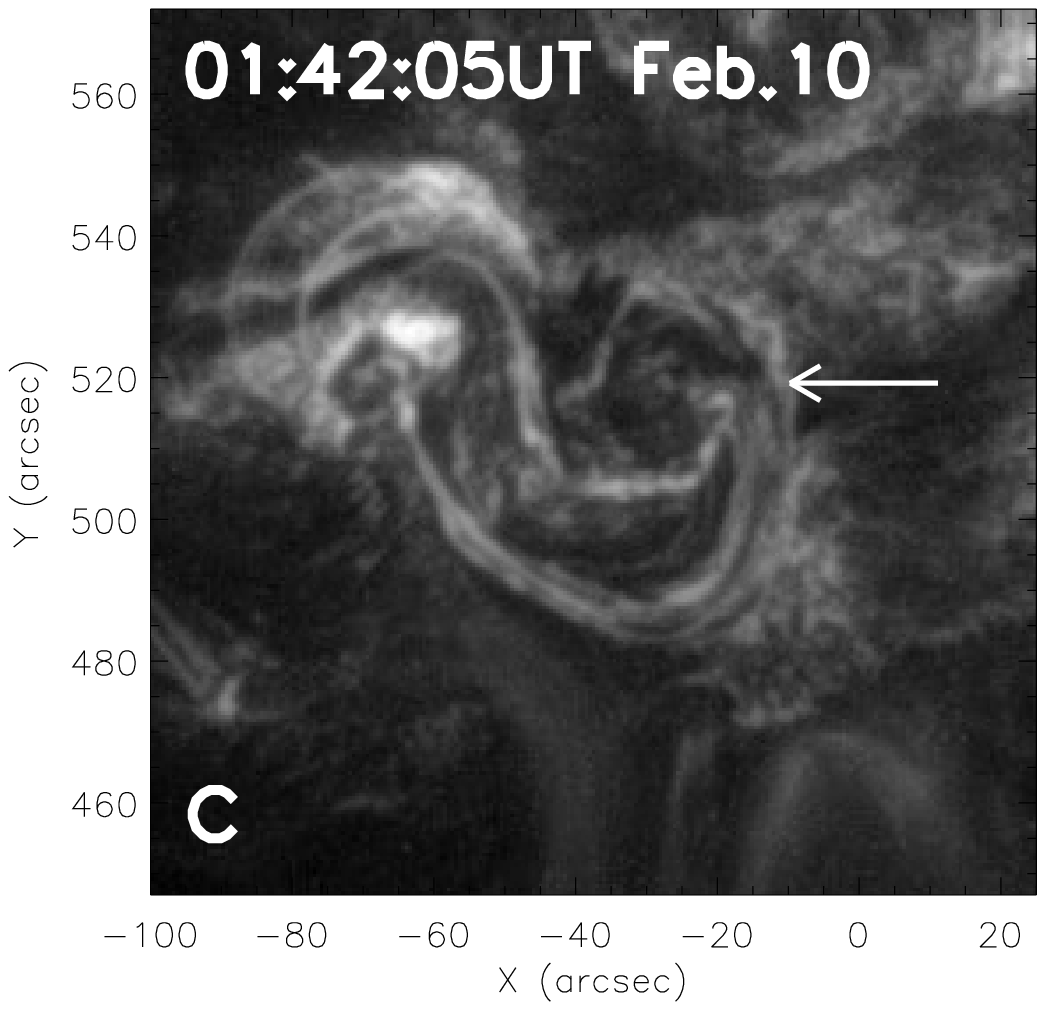}%
    \includegraphics[width=4cm]{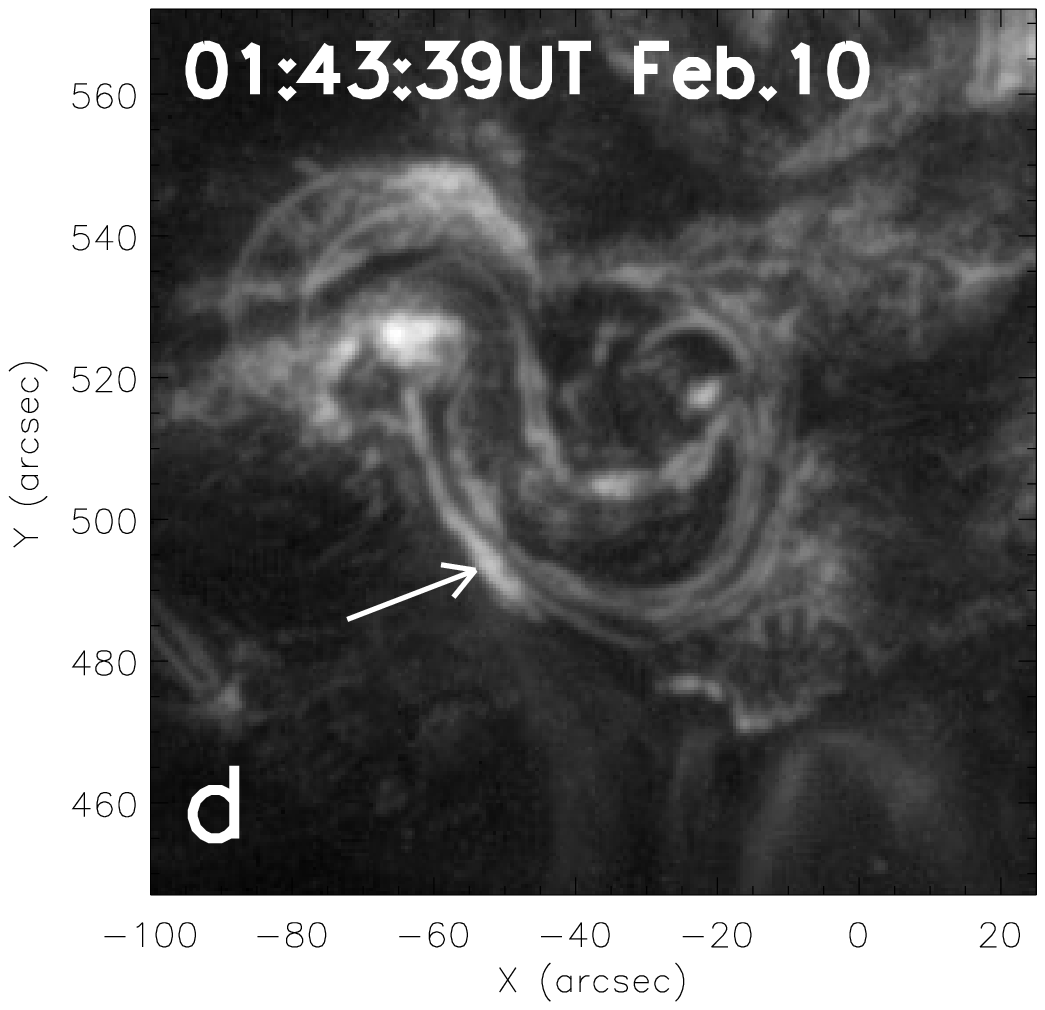}\\
   \includegraphics[width=4cm]{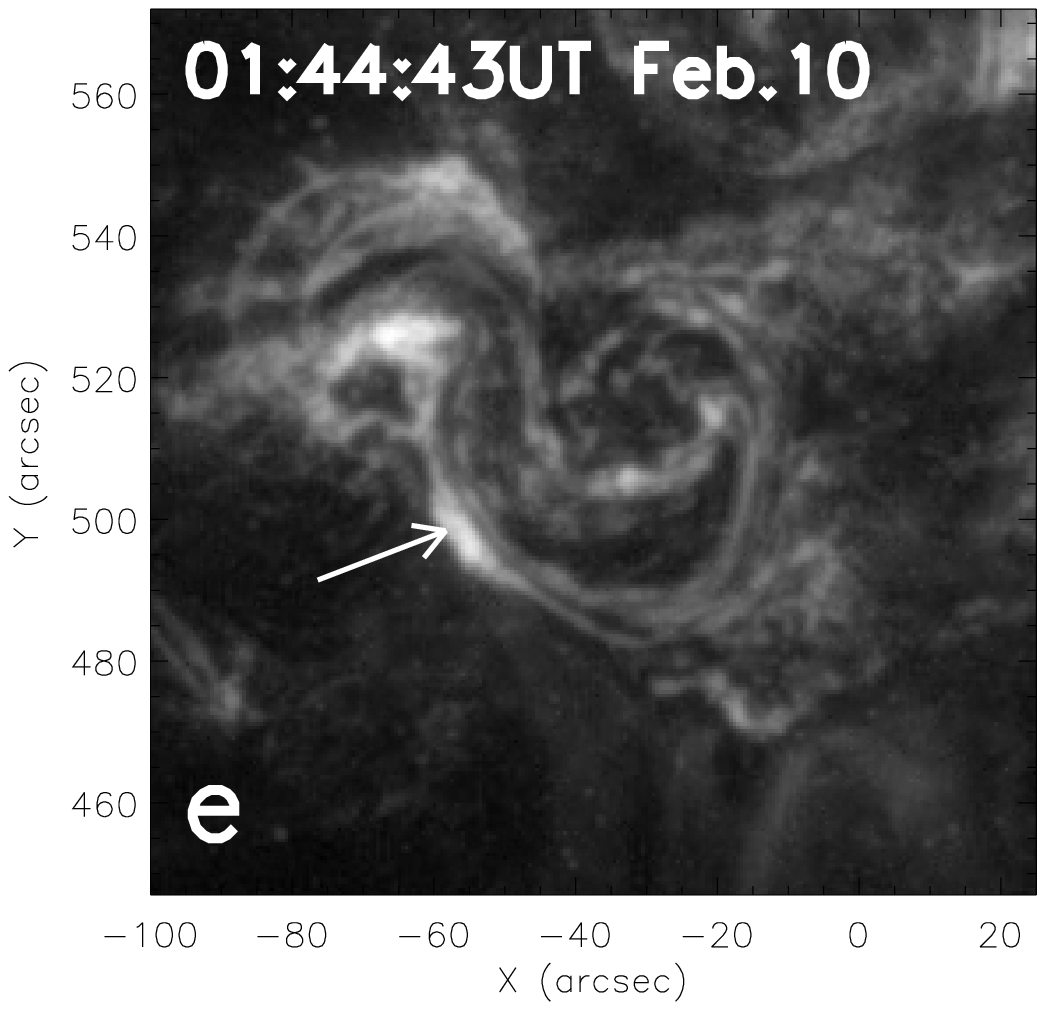}%
   \includegraphics[width=4cm]{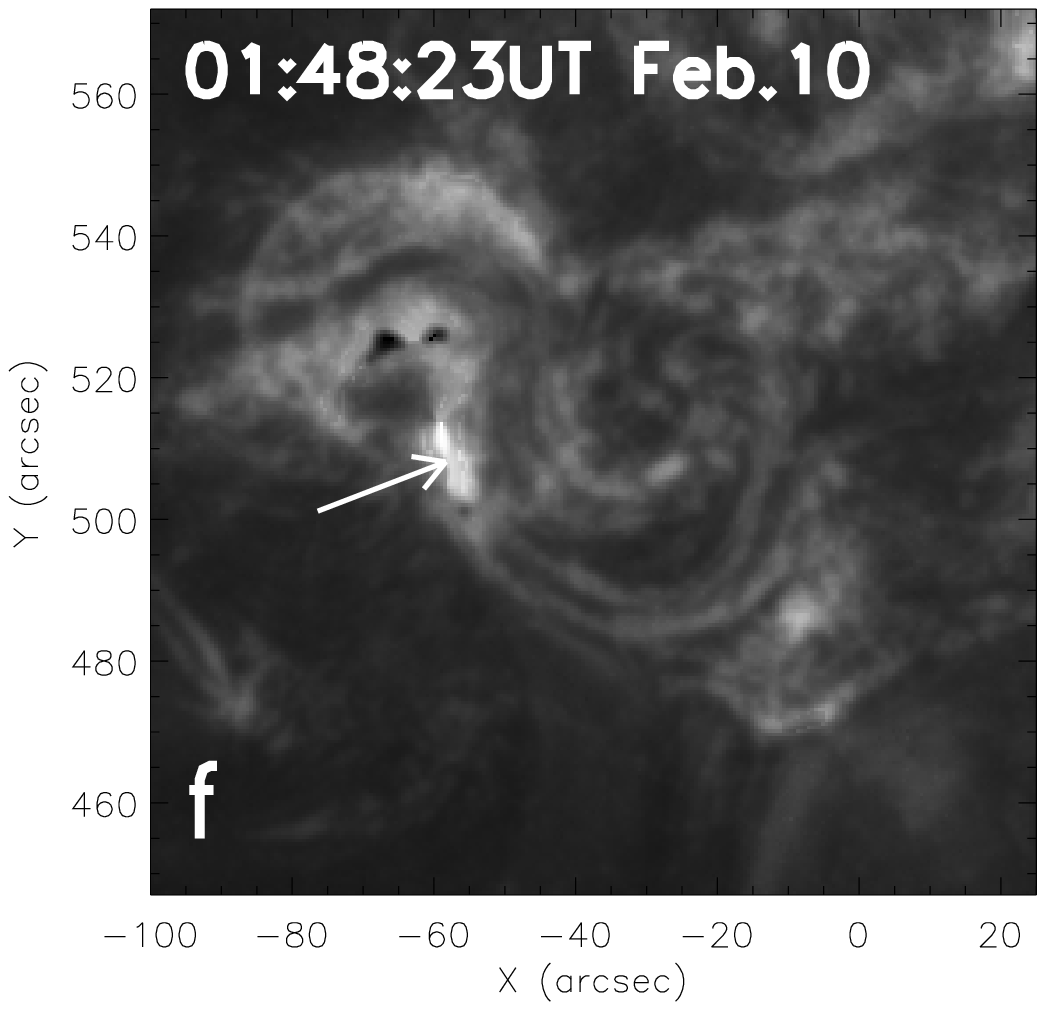}%
    \includegraphics[width=4cm]{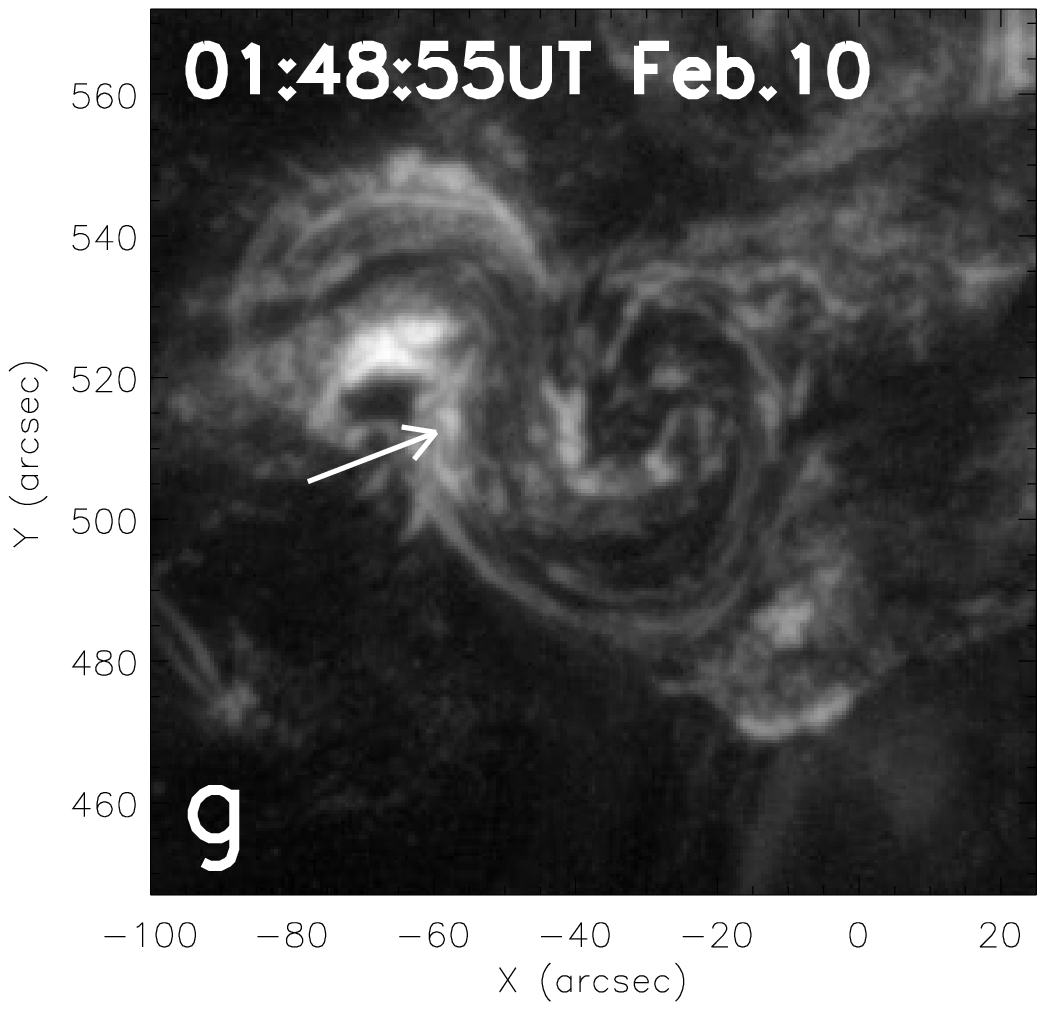}%
    \includegraphics[width=4cm]{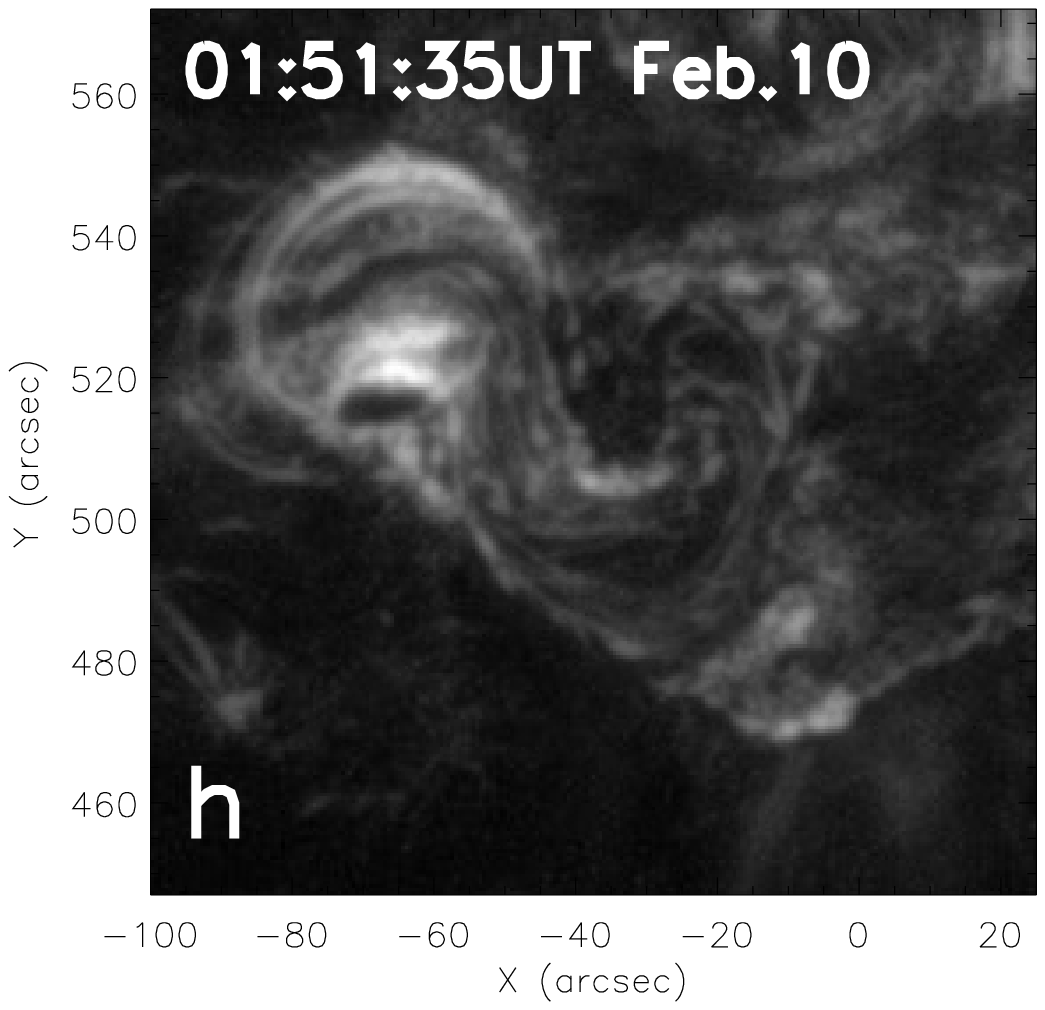}\\
    \includegraphics[width=4cm]{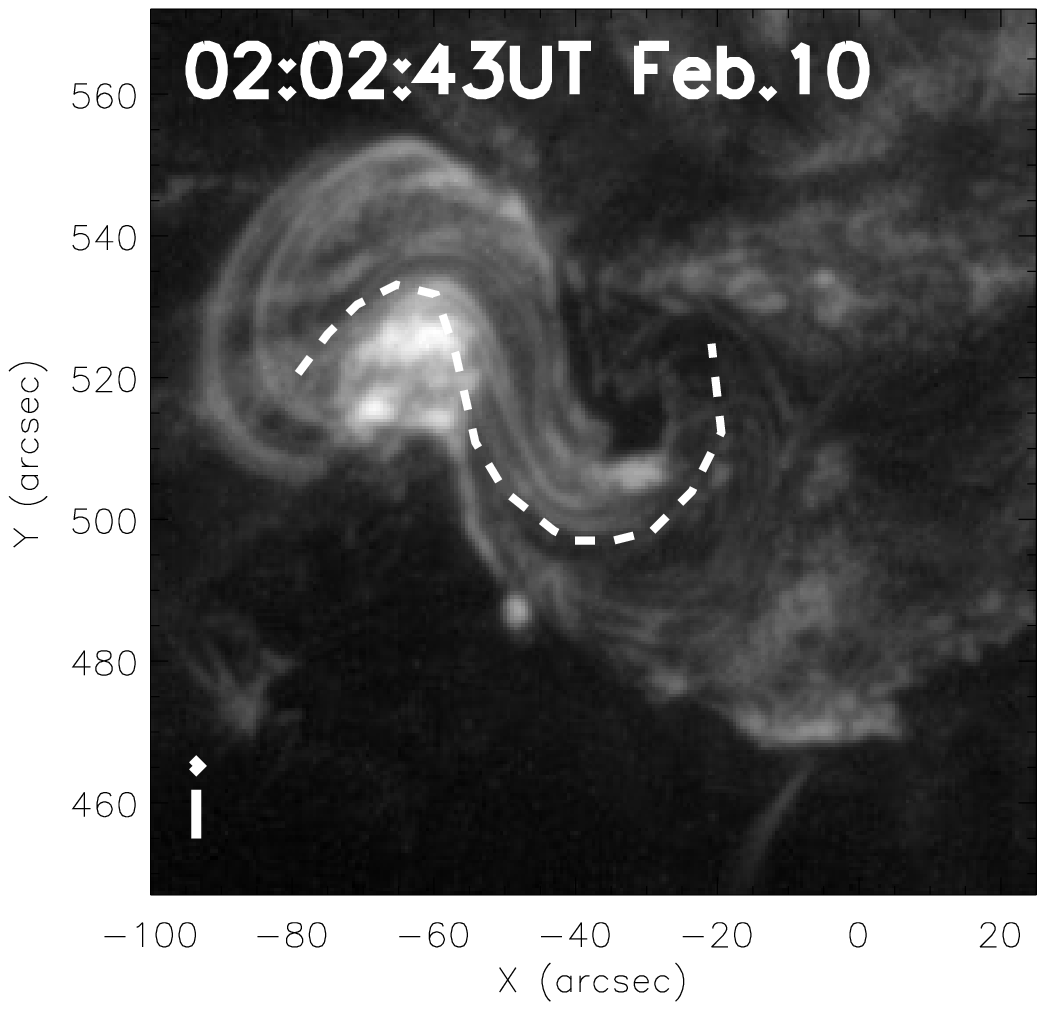}%
   \includegraphics[width=4cm]{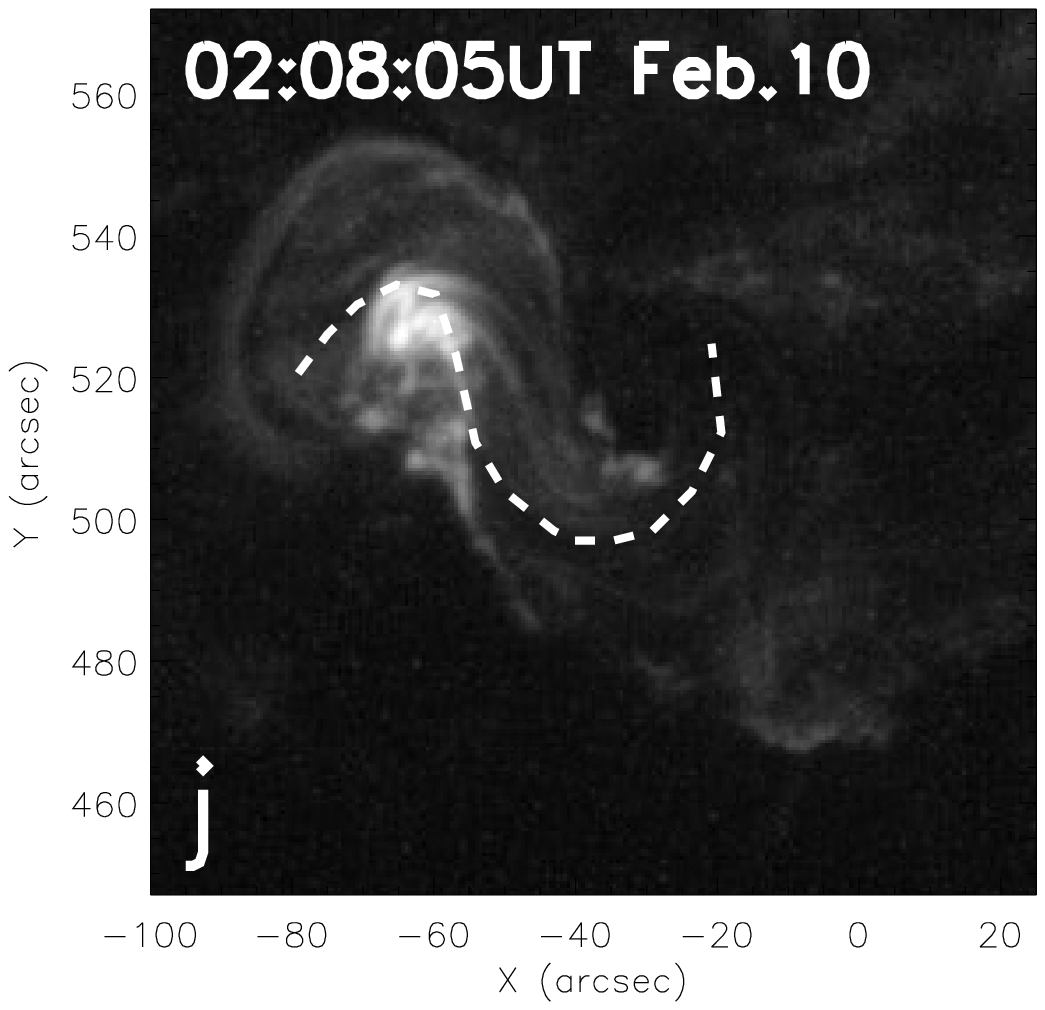}%
    \includegraphics[width=4cm]{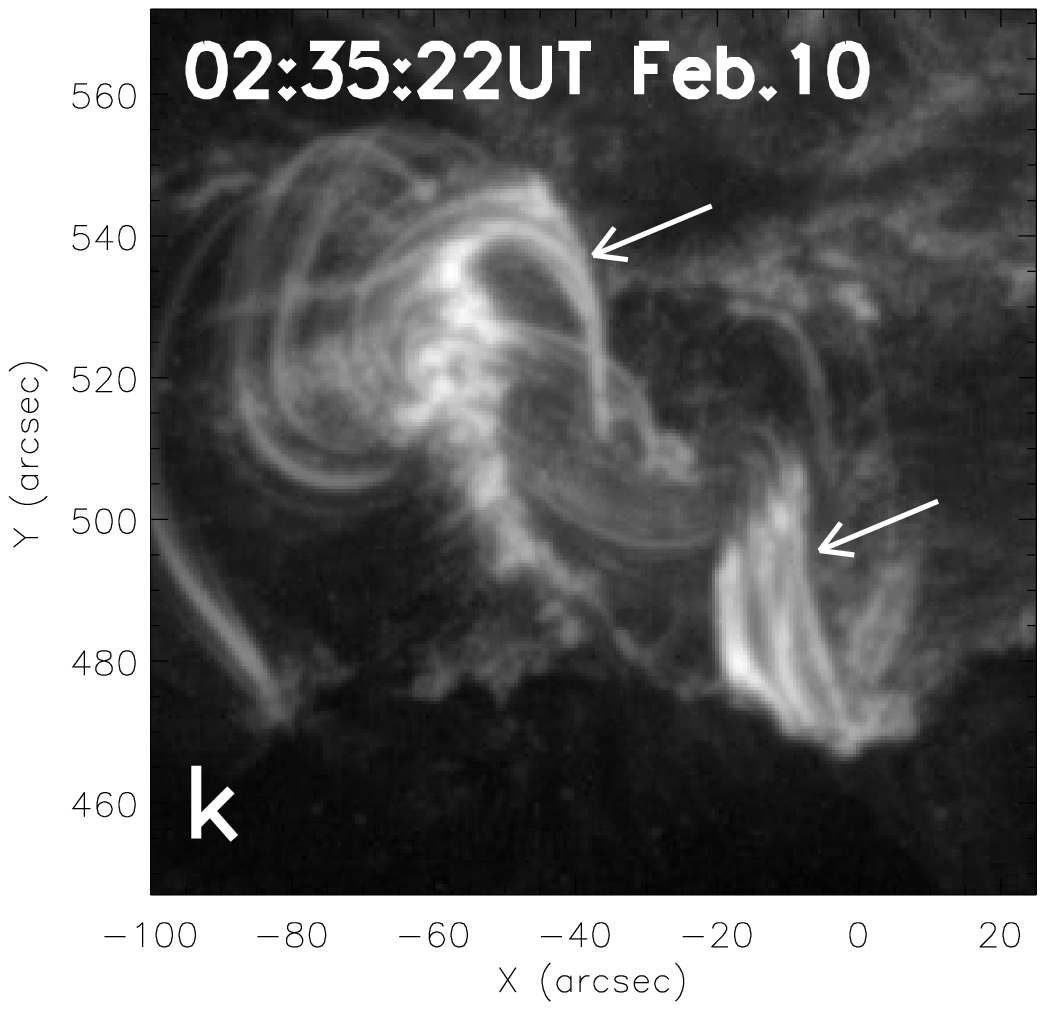}%
    \includegraphics[width=4cm]{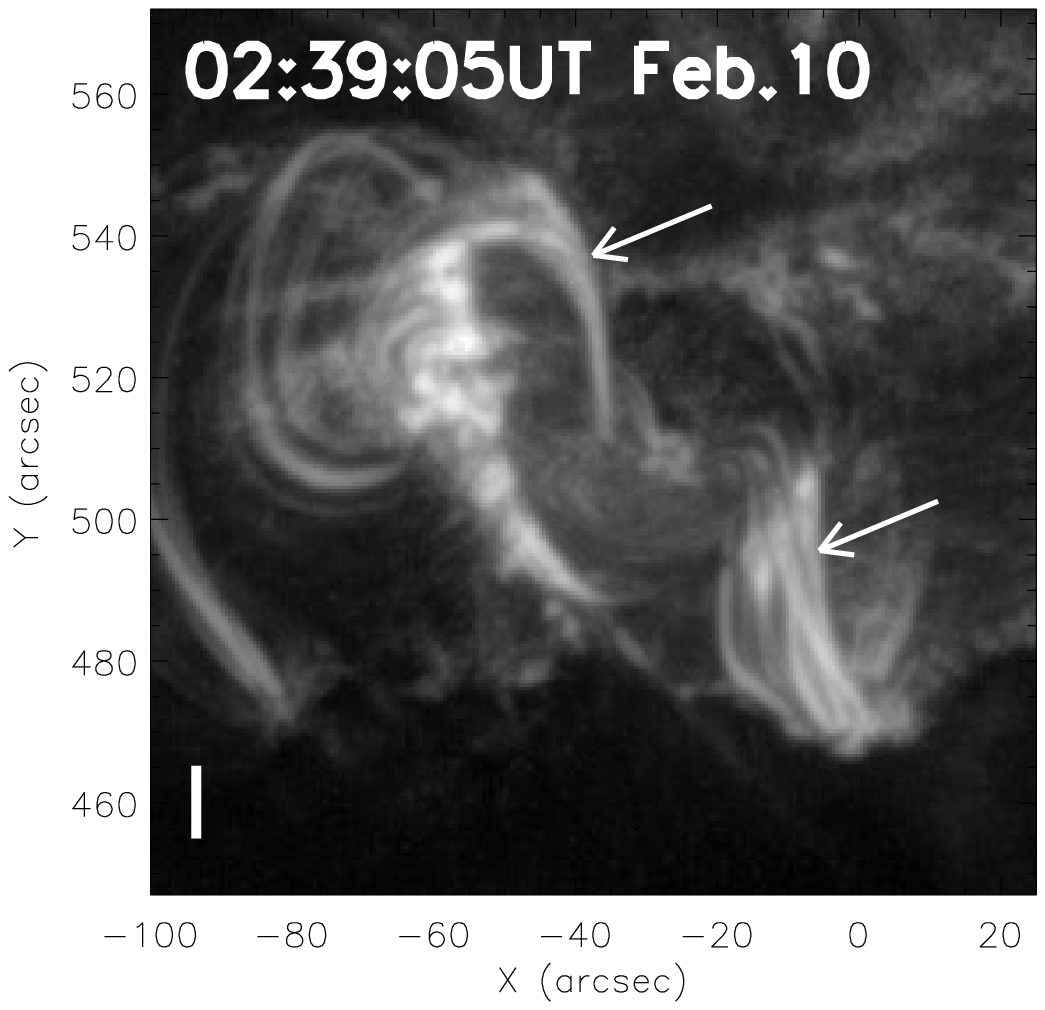}\\
\caption{A sequence of 171 \AA\ images showing the second successful eruption of the filament from 01:38:42UT to 02:39:05UT on February
10, 2000. The dashed lines denote the inverse S-shaped filament. The arrows are described in the text.}%% no full stop at the end
\end{figure*}

\begin{figure*}
\centering
   \includegraphics[width=12cm]{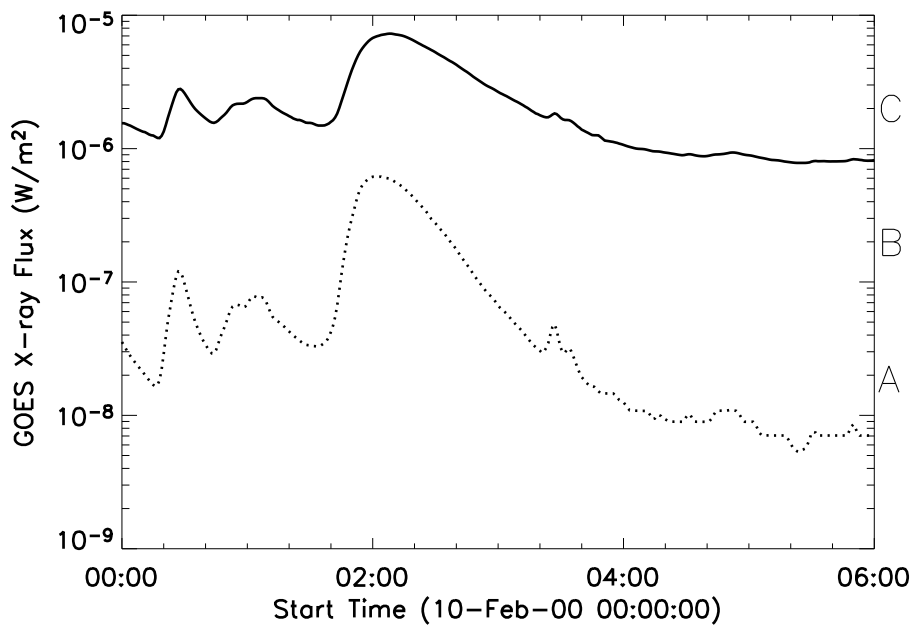}
\caption{The evolution of GOES soft X-ray emission for the C7.3 flare on February
10, 2000 (Solid line: 1-8 \AA; Dashed line: 0.5-4 \AA).}%% no full stop at the end
\end{figure*}

\begin{figure*}
\centering
   \includegraphics[width=4cm]{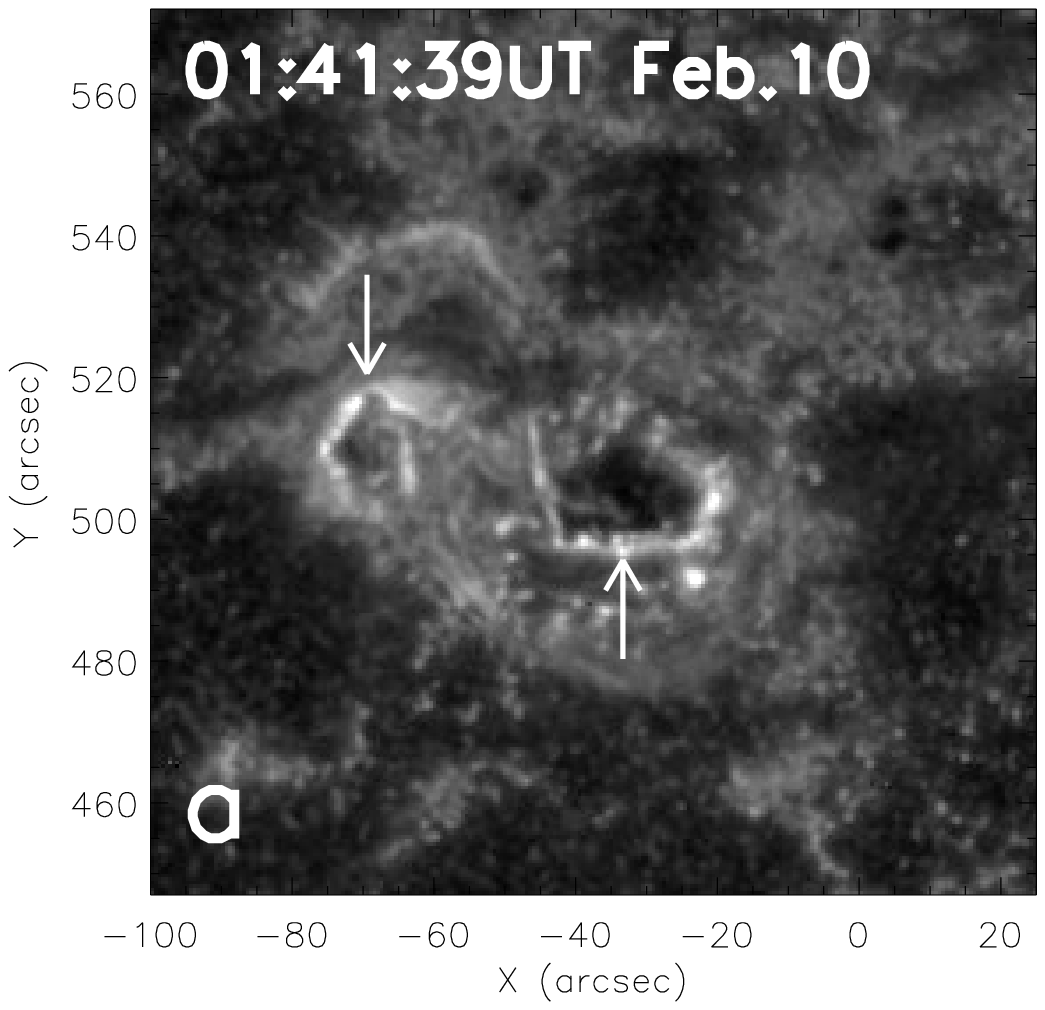}%
   \includegraphics[width=4cm]{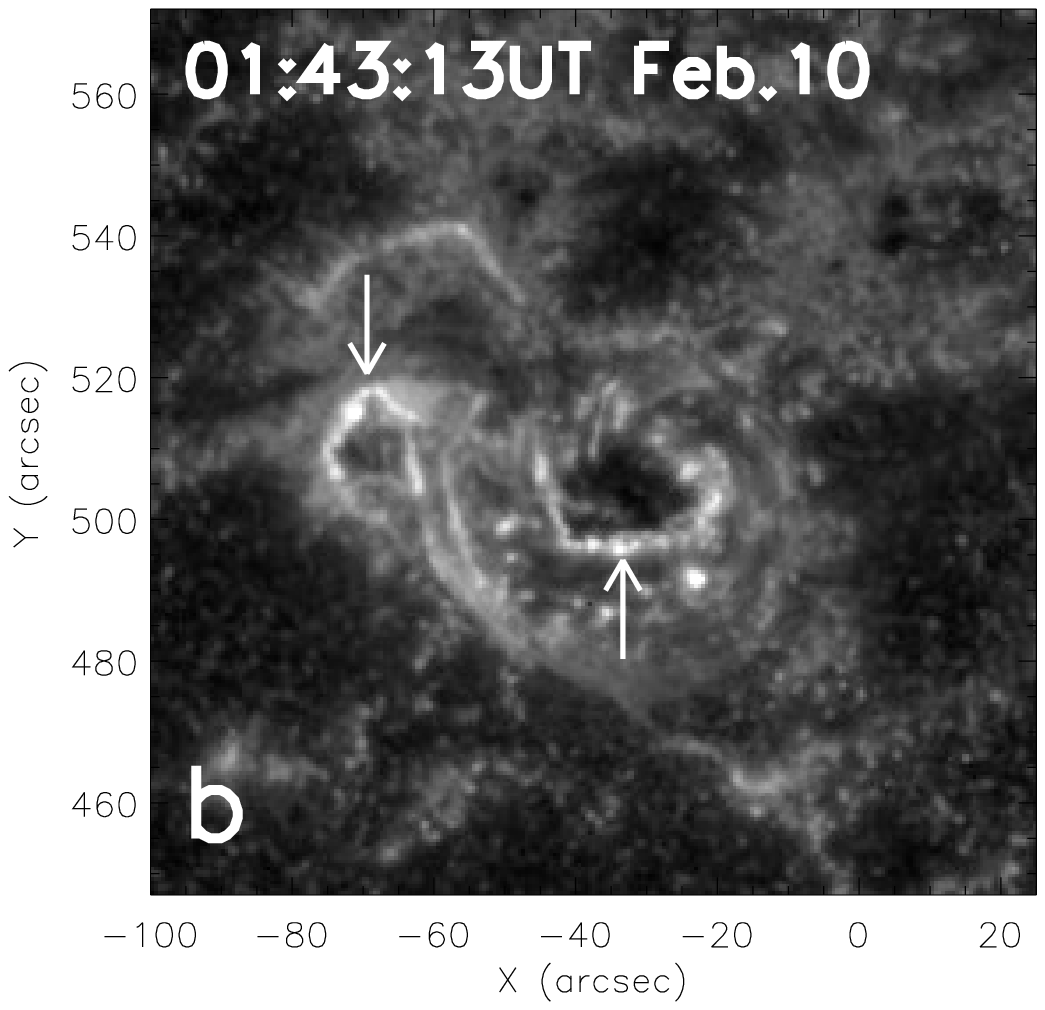}%
    \includegraphics[width=4cm]{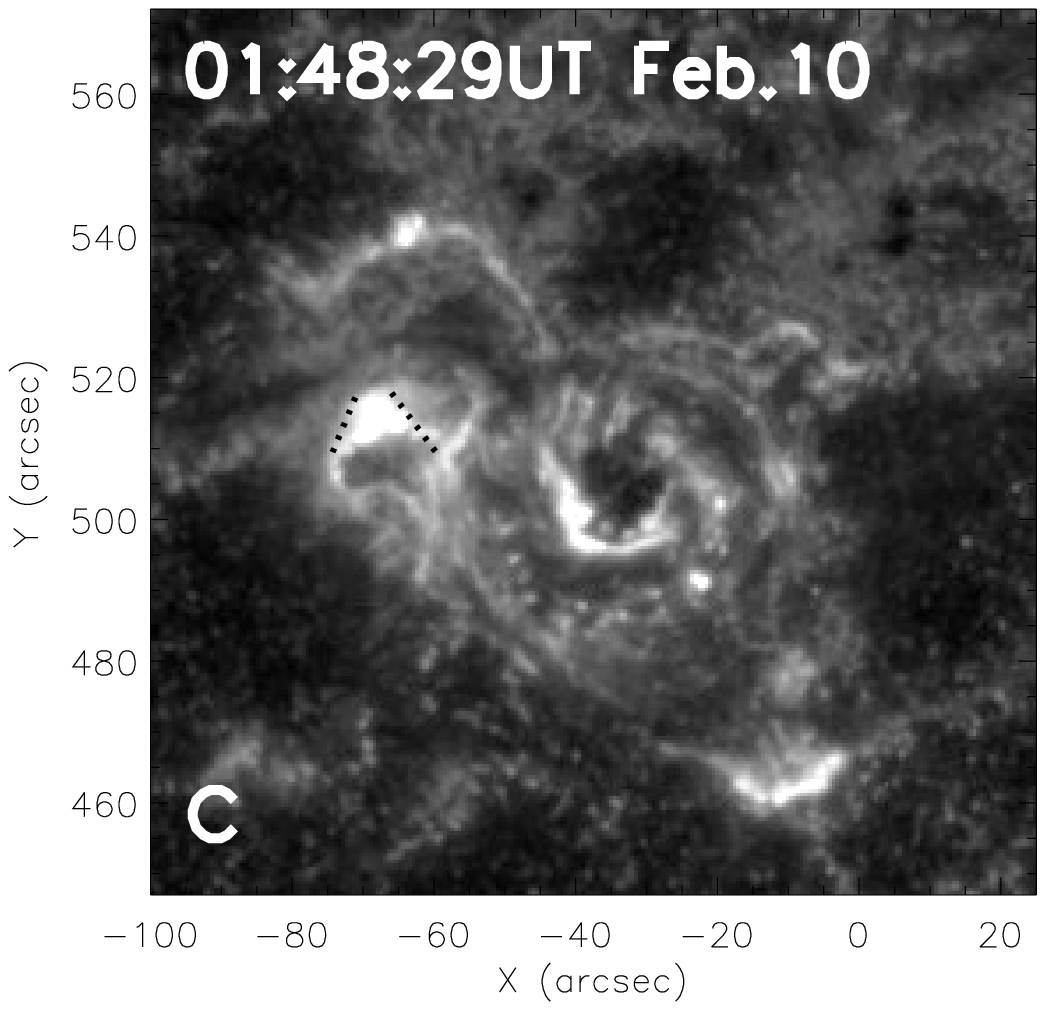}%
    \includegraphics[width=4cm]{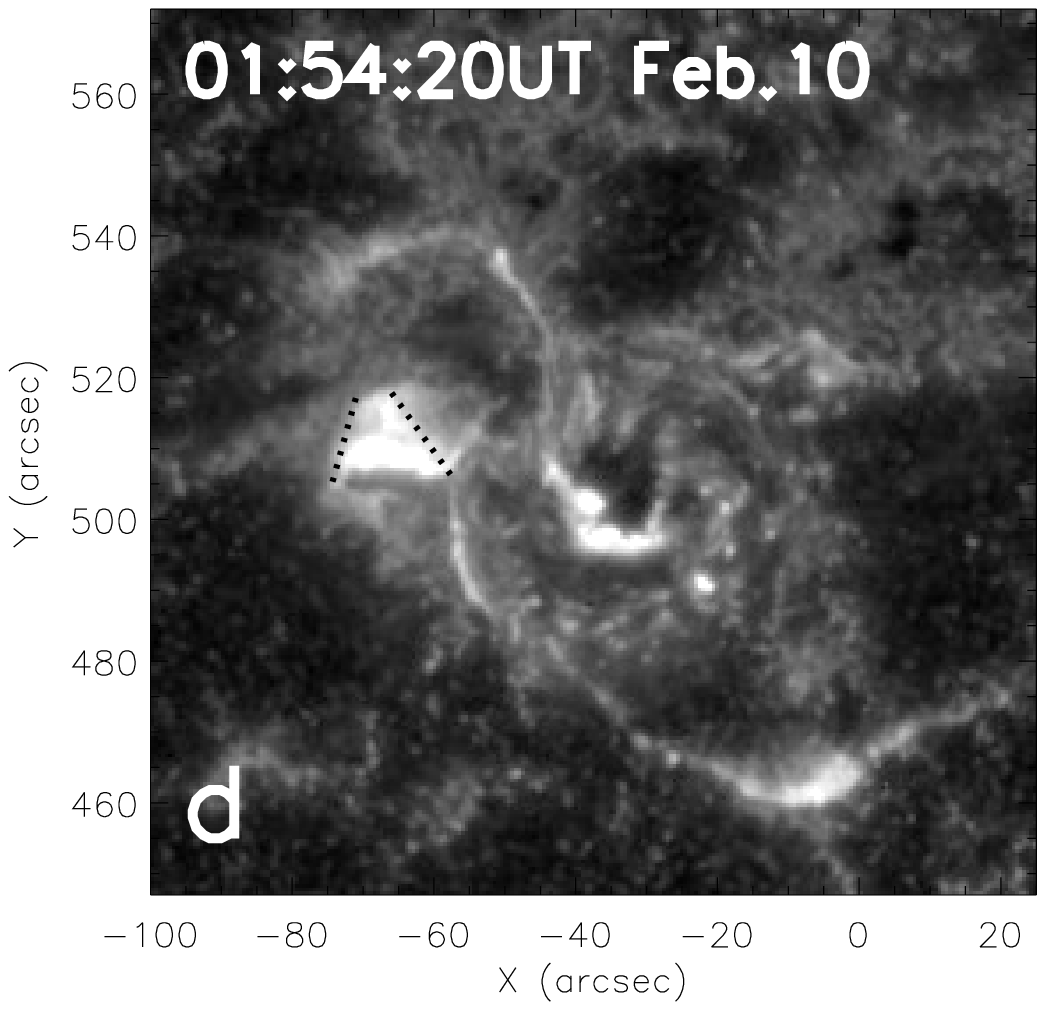}\\
    \includegraphics[width=4cm]{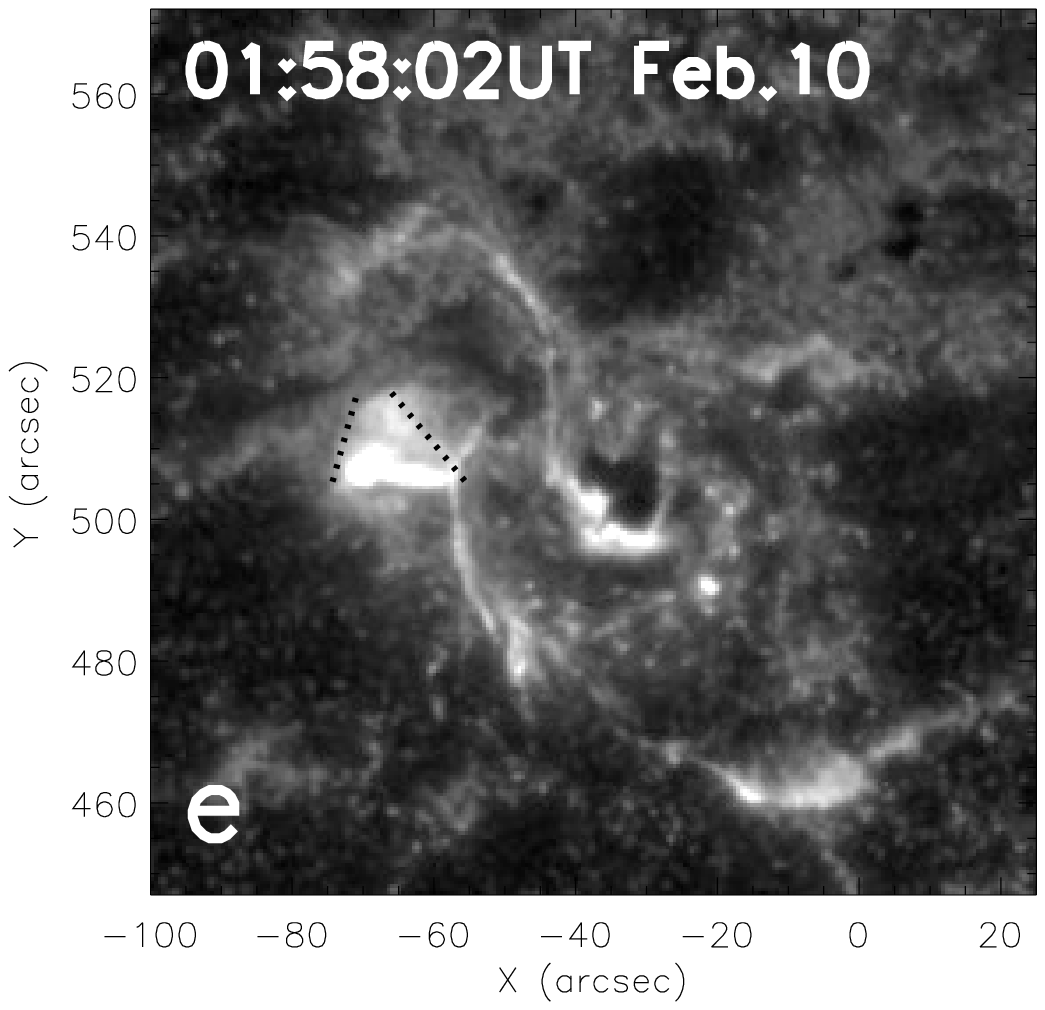}%
   \includegraphics[width=4cm]{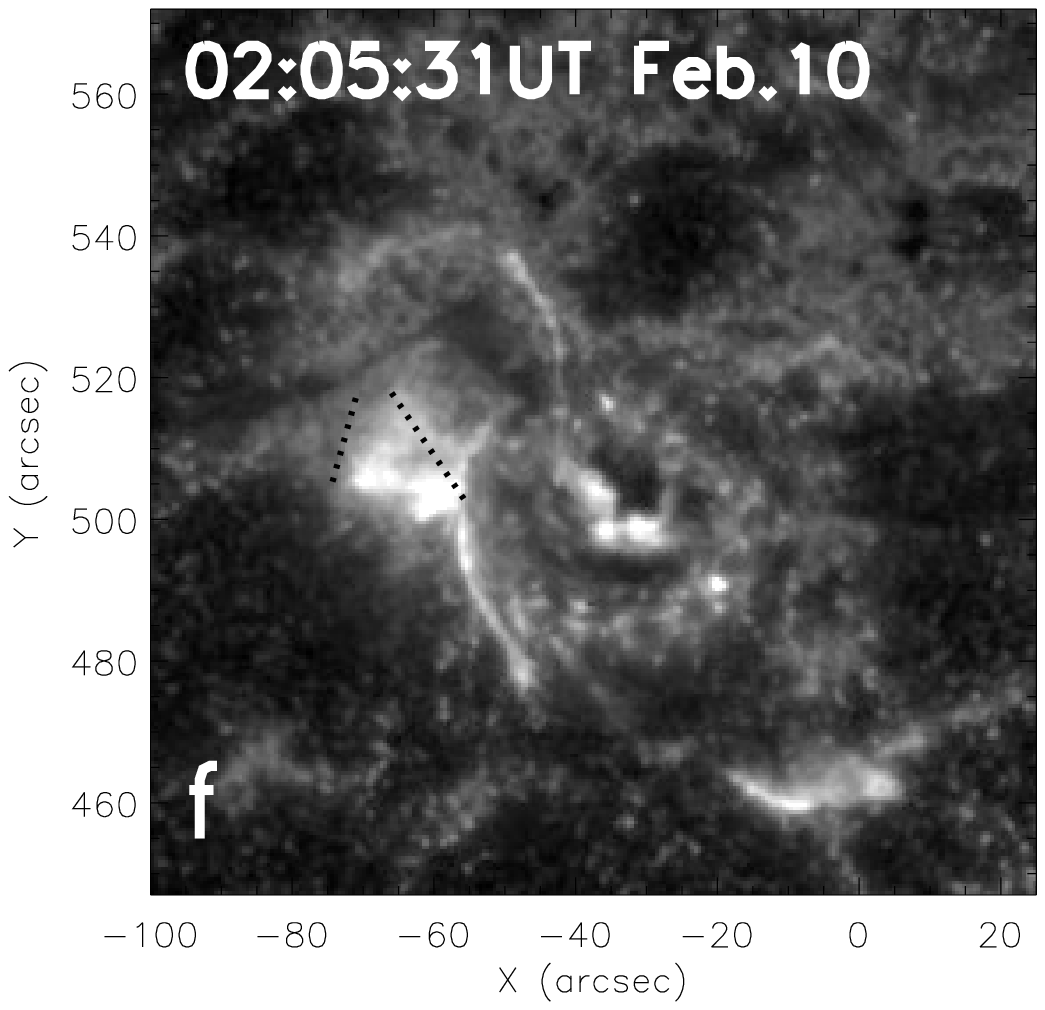}%
    \includegraphics[width=4cm]{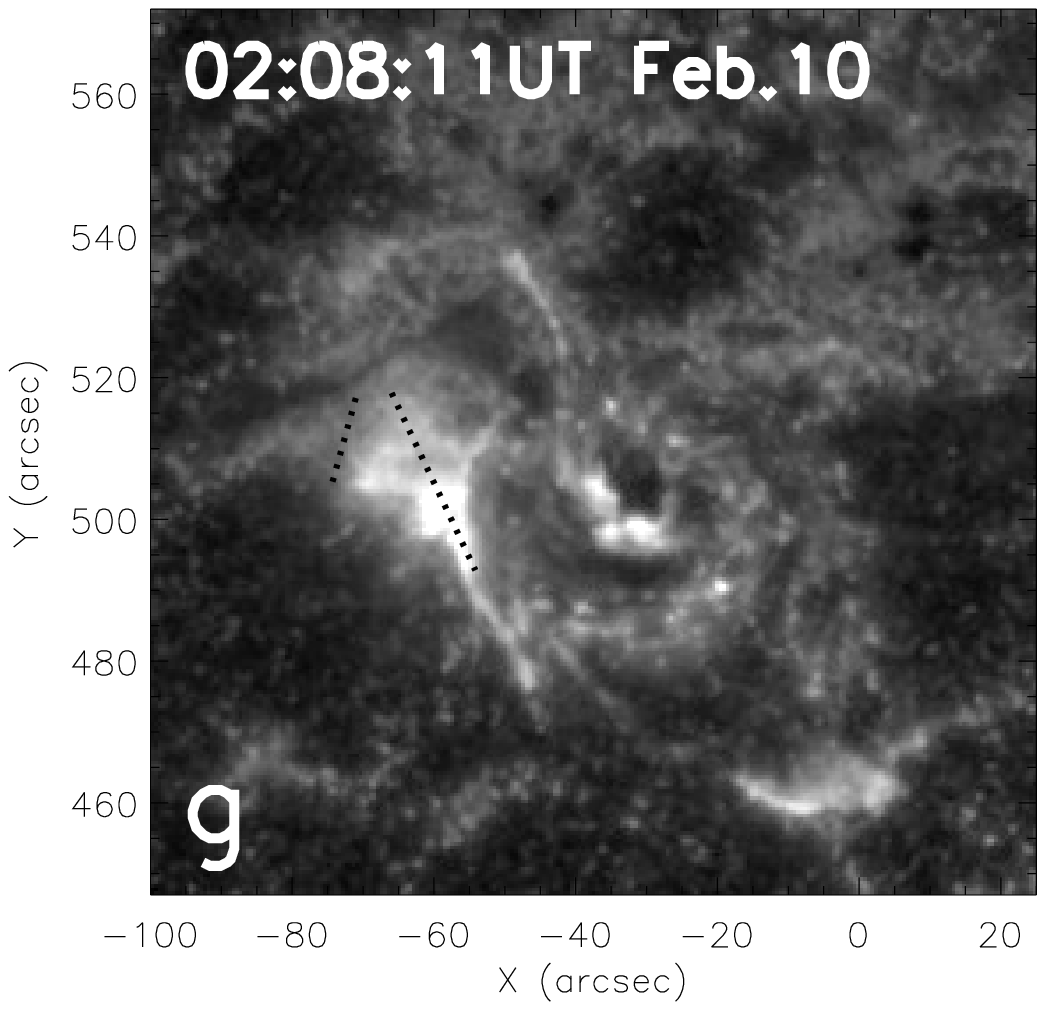}%
    \includegraphics[width=4cm]{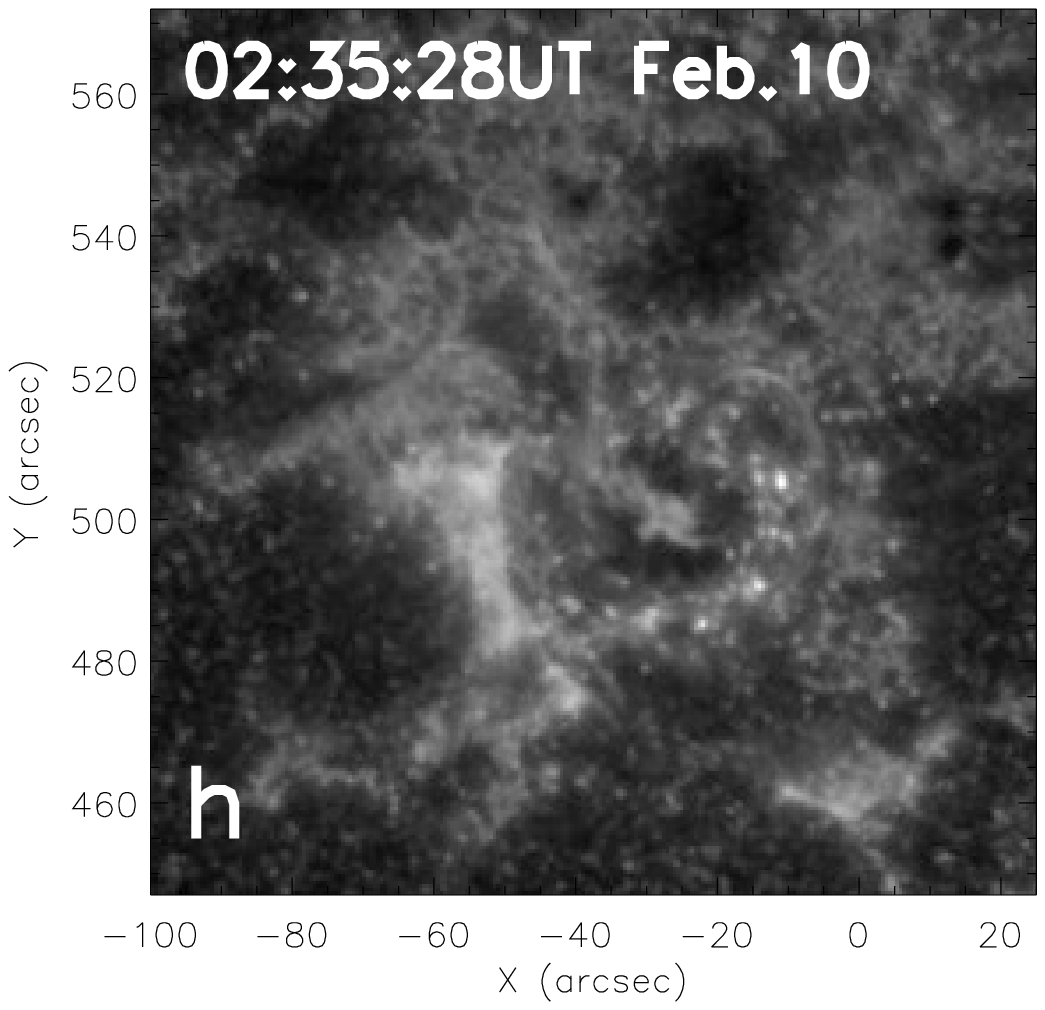}\\
\caption{The evolution of the flare ribbons from 01:41:39UT to 02:35:28UT on Febuary 10, 2000. The two arrows in Figs. 9a and 9b indicate the two flare ribbons. The dotted lines denote the direction of the left flare ribbon expansion from the north of the following sunspot to the southwest.}
\end{figure*}

%\bibitem[Li et al.(1998)]{li98} Li, K.J., Schmieder, B., Li, Q. 2003, \aaps, 131, 99

%\bibitem[White et al.(1977)]{whi77} White, O. R., Trotter, Do. E. 1977, \apjs, 33, 391

%\begin{figure*}
%  \centering
%   \includegraphics[width=15cm]{fig1.eps}
%\caption{The distribution of the monthly numbers of four type flares. The blue, red, black, and green lines indicate the numbers of B, C, M, X flare from 1975 September %to 2008 May respectively.}
 %       \label{}
%   \end{figure*}


\begin{thebibliography}{}
\bibitem[Amari, T.(2000)]{ama00} Amari, T., Luciani, J. F., Mikic, Z., Linker, J. 2000, \apj, 529, 49

\bibitem[Aulanier, G.(2010)]{ama00} Aulanier, G., T$\ddot{o}$r$\ddot{o}$k, T., D$\acute{e}$moulin, P., DeLuca, E. E. 2010, \apj, 708, 314

\bibitem[Barnes, C. W.(1972)]{bar72} Barnes, C. W., Sturrock, P. A. 1972, \apj, 174, 659

\bibitem[Brown, D.(2003)]{bro03} Brown D. S., Nightingale R. W., Alexander D., Schrijver C. J., Metcalf
T. R., Shine R. A., Title A. M., Wolfson C. J. 2003, \solphys, 216, 79

\bibitem[Brueckner et al. (1995)]{bru95} Brueckner, G. E., Howard, R. A., Koomen, M. J., Korendyke, C. M., Michels, D. J., Moses, J. D. et al. 1995, \solphys, 162, 357

\bibitem[Bi, Y. (2011)]{bi11} Bi, Y., Jiang, Y. C., Yang, L. H., Zheng, R. S. 2011, \na, 16, 276

\bibitem[Canfield et al. (1999)]{can99} Canfield R. C., Hudson H. S., McKenzie D. E., 1999, Geophys. Res. Lett., 26, 627

\bibitem[Canfield et al. (2007)]{can07} Canfield, Richard C., Kazachenko, M. D., Acton, L. W., Mackay, D. H., Son, J., Freeman, T. L., 2007, \apj, 671, L81

\bibitem[Domingo et al.(1995)]{dom95} Domingo, V., Fleck, B., \& Poland, A. I. 1995, \solphys, 162, 1

\bibitem[Evershed, J.(1910)]{eve10} Evershed, J., 1910, \mnras, 70, 217

\bibitem[Fan, Y.(2001)]{fan01} Fan, Y. 2001, \apj, 554, L111

\bibitem[Fan, Y.(2009)]{fan09} Fan, Y. 2009, \apj, 697, 1529

\bibitem[Gopasyuk, S.(1965)]{gop65} Gopasyuk, S. I., 1965, Izv. Krym. Astrofiz. Obs., 33, 100

\bibitem[Gibson, S. E.(2002)]{gib02} Gibson, S. E., Fletcher, L., Del Zanna, G., Pike, C. D., Mason, H. E., Mandrini, C. H., D$\acute{e}$moulin, P., Gilbert, H., Burkepile, J., Holzer, T., et al. 2002, \apj, 574, 1021

\bibitem[Gibson, S. E.(2004)]{gib04} Gibson, S. E., Fan, Y., Mandrini, C., Fisher, G., Demoulin, P. 2004, \apj, 617, 600

\bibitem[Green, L. M.(2009)]{gre09} Green, L. M., Kliem, B. 2009, \apj, 700, L83

\bibitem[Green, L. M.(2011)]{gre11} Green, L. M., Kliem, B., Wallace, A. J. 2011, \aap, 526, 2

\bibitem[Hood, A. W.(1991)]{hoo91} Hood, A. W. 1991, Solar System Magnetohydrodynamics, ed. E. R. Priest \& A. W. Hood (London: Cambridge Univ. Press), 307

\bibitem[Hudson, H. S.(1998)]{hud98} Hudson, H. S., Lemen, J. R., St. Cyr, O. C., Sterling, A. C., Webb, D. F.
1998, Geophys. Res. Lett., 25, 2481

\bibitem[Handy, B.N.(1999)]{han99} Handy, B.N., Acton, L.W., Kankelborg, C.C.,Wolfson, C.J.,Akin, D.J., Bruner, M.E.,Caravalho, R., Catura, R.C., et al. 1999, \solphys, 187, 229

\bibitem[Jiang, Y.(2007)]{jia07} Jiang, Y.C., Chen, H.D., Shen, Y.D., Yang, L.H., Li, K.J. 2007, \solphys, 240, 77

\bibitem[Jiang, Y.(2012)]{jia12} Jiang, Y.C., Zheng, R.S., Yang, J.Y., Hong, J.C., Yi, B., Yang, D. 2012, \apj, 744, 50

\bibitem[Jing, J.(2004)]{jin04} Jing, J., Yurchyshyn, V. B., Yang, G., Xu, Y., Wang, H. 2004, \apj, 614, 1054

\bibitem[Kliem, B.(2004)]{kli04} Kliem, B., Titov, V. S., T$\ddot{o}$r$\ddot{o}$k, T. 2004, \aap, 413, L23

\bibitem[Kazachenko, M. D.(2009)]{kaz99} Kazachenko, M. D., Canfield, R. C., Longcope, D. W., Qiu, J., Des Jardins, A., Nightingale, R. W. 2009, \apj, 704, 1146

\bibitem[Kumar, P.(2010)]{kum10} Kumar, P., Manoharan, P. K., \& Uddin, W. 2010, \apj, 710, 1195

\bibitem[Liu, Y.(2002)]{liu02} Liu, Y., Zhao, X.P., Hoeksema, J.T., Scherrer, P.H., Wang, J., Yan, Y. 2002, \solphys, 206, 333

\bibitem[Low, B.C.(2003)]{low03} Low, B.C. \& Berger, M.A. 2003, \apj, 589, 644

\bibitem[Leamon, R.J.(2003)]{lea03} Leamon, R.J., Canfield, R.C., Blehm, Z., Pevtsov, A.A. 2003, \apj, 596, L255

\bibitem[Liu, C.(2007)]{liu07} Liu, C., Lee, J., Yurchyshyn, V., Deng, N., Cho, K. 2007, \apj, 669, 1372

\bibitem[Li, L.P.(2009)]{li09} Li, L.P., \& Zhang, J. 2009, \apj, 706, L17

\bibitem[Maltby, P.(1964)]{mal64} Maltby, P., 1964, Astrophys. Nor., 8, 205

\bibitem[Magara, T.(2001)]{mga01} Magara, T. \& Longcope, D. W. 2001, \apj, 559, 55

\bibitem[Moore, R. L.(2001)]{moo01} Moore, R. L., Sterling, A. C., Hudson, H. S., Lemen, J. R. 2001, \apj, 552,
833

\bibitem[McKenzie, D. E.(2008)]{mck08} McKenzie, D. E., \& Canfield, R. C. 2008, \aap, 481, L65

\bibitem[Min, S.(2009)]{min09} Min, S. \& Chae, J. 2009, \solphys, 258, 203

\bibitem[Nightingale, R. W.(2002)]{nig02} Nightingale, R. W., Brown, D. S., Metcalf, T. R., et al. 2002, in Yohkoh 10th Anniv. Meeting, Muti-wavelength
Observattions of Coronal Structure and Dynamics, ed. P. C. H. Martens, \& D. Cauffman, 149

\bibitem[Pevtsov, A. A.(2002)]{pev02} Pevtsov, A. A. 2002, \solphys, 207, 111

\bibitem[Park, S.H.(2010)]{par10} Park, S.H., Chae, J., Jing, J., Tan, C. Wang, H. 2010, \apj, 720, 1102

\bibitem[Rust, D.M. (1994)]{rus94} Rust, D.M., Sakurai, T., Gaizauskas, V., Hofmann, A., Martin, S.M., Priest, E.R., Wang, J., 1994, \solphys, 153, 1

\bibitem[Regnier, S.(2006)]{reg06} R$\acute{e}$gnier, S., Canfield R. C., 2006, \aap, 451, 319

\bibitem[Regnier, S.(2004)]{reg04} R$\acute{e}$gnier, S., Amari T., 2004, \aap, 425, 345

\bibitem[Ravindra, B.(2011)]{rav11} Ravindra, B., Yoshimura, K., Dasso, S. 2011, \apj, 743, 33

\bibitem[Stenflo, J. O.(1969)]{ste69} Stenflo, J. O. 1969, \solphys, 8, 115

\bibitem[Scherrer, P. H.(1995)]{sch95} Scherrer, P. H., Bogart, R.S., Bush, R.I., Hoeksema, J.T., Kosovichev, A.G., Schou, J., Rosenberg, W., Springer, L., et al. 1995, \solphys, 162, 169

\bibitem[Sterling, A. C.(1997)]{ste97} Sterling, A. C., \& Hudson, H. S. 1997, \apj, 491, L55

\bibitem[Sterling, A. C.(2000)]{ste00} Sterling, A. C., Hudson, H. S., Thompson, B. J., Zarro, D. M. 2000, \apj, 532, 628

\bibitem[Schrijver, C. J.(2008)]{sch08} Schrijver, C. J., Elmore, C., Kliem, B., T$\ddot{o}$r$\ddot{o}$k, T., Title, A. M. 2008, \apj, 674, 586

\bibitem[Savcheva, A.(2009)]{sav09} Savcheva, A., \& van Ballegooijen, A. 2009, \apj, 703, 1766

\bibitem[Titov, V.S.(1999)]{tit99} Titov, V.S., \& D$\acute{e}$moulin, P. 1999, \aap, 351, 707

\bibitem[T$\ddot{o}$r$\ddot{o}$k(2003)]{tor03} T$\ddot{o}$r$\ddot{o}$k, T., Kliem, B. 2003, \aap, 406, 1043

\bibitem[Tian, L.(2006)]{tia06} Tian, L., Alexander, D., 2006, \solphys, 233, 29

\bibitem[Tripathi, D.(2009)]{tri09} Tripathi, D., Kliem, B., Mason, H. E., Young, P.r R., Green, L. M. 2009, \apj, 698, L27

\bibitem[Vrsnak, B.(1991)]{vrs91} $Vr\check{s}nak$, $Ru\check{z}djak$, \& Rompolt, B. 1991, \solphys, 136, 151

\bibitem[Wang, H.(2007)]{wan07} Wang, H., Liu, C., Jing, J., Yurchyshyn, V. 2007, \apj, 671, 973

\bibitem[Yan, X.L.(2007)]{yan07} Yan X. L., Qu Z. Q., 2007, \aap, 468, 1083

\bibitem[Yan, X.L.(2008)]{yan08} Yan X. L., Qu Z. Q., Xu C. L., 2008, \apj, 682, L65

\bibitem[Yan, X.L.(2008)]{yan08} Yan, X.L., Qu, Z.Q., Kong, D.F. 2008, \mnras, 391, 1887

\bibitem[Yan, X.L.(2009)]{yan09} Yan, X.L., Qu, Z.Q., Xu, C.L., Xue, Z.K., Kong, D.F. 2009, RAA, 9, 596

\bibitem[Yan, X.L.(2011)]{yan11} Yan, X.L., Qu, Z.Q., Kong, D.F. 2011, \mnras, 414, 2803

\bibitem[Zhao, J. W.(2003)]{zha03} Zhao, J. W., \& Kosovichev, A. G. 2003, \apj, 591, 446

\bibitem[Zhang, M.(2006)]{zha06} Zhang, M., Flyer, N., \& Low, B. C. 2006, \apj, 644, 575

\bibitem[Zhang, J.(2007)]{zha07} Zhang J., Li L. P., Song Q., 2007, \apj, 662, L35

\bibitem[Zhang, Y.(2008)]{zha08} Zhang, Y., Liu, J., Zhang, H. 2008, \solphys, 247, 39
\end{thebibliography}
\end{document}